\documentclass[%
 reprint,
 amsmath,amssymb,
 prb,
]{revtex4-1}
\usepackage{graphicx}
\usepackage{dcolumn}
\usepackage{bm}

\begin{document}

\preprint{APS/123-QED}

\preprint{APS/123-QED}
\title{The competition between the intrinsic and Rashba spin-orbit coupling and effects of correlations on Rashba SOC-driven transitions in the Kane-Mele model} 
\author{Tao Du}
\email{Corresponding author: dutao@ymu.edu.cn}
\author{Yuexun Li}
\author{Helin Lu}
\author{Hui Zhang}
\affiliation{Department of Physics, Yunnan Minzu University, Kunming 650504, P. R. China}
\author{Song Du}
\email{Corresponding author: songduscau@163.com}
\affiliation{College of Science, Sichuan Agricultural University, Ya'an 625014, P. R. China}
\date{\today}

\begin{abstract}
We investigate, firstly, the effects of the Rashba SOC on the band structrue of the Kane-Mele model. The competition between the Rashba SOC and the intrinsic SOC can lead to the rich phenomenology. The Rashba SOC can drive the indirect and direct energy gap to close successively, but maintain the band touching between the valence band and the conduction band when the Rashba SOC is large enough to dominant the competition. We find that these touching points are located at $K$ and $K^{\prime}$ or/and some $2\pi/3$ rotationally symmetric points around $K$ and $K^{\prime}$ in the Brillouin zone. The indirect and direct energy gap closings correspond to the topologically trivial and non-trivial phase transitions respectively. For the small intrinsic SOC, the topologically non-trivial transition occurs when the ratio of the Rashba SOC to the intrinsic SOC is equal to the classical result, i.e. $2\sqrt{3}$. For the large intrinsic SOC, however, we find that the ratio decreases with the increasing intrinsic SOC. Secondly, using the slave-rotor mean field method we investigate the influences of the correlation on the Rashba SOC-driven topologically trivial and non-trivial transition in both the charge condensate and Mott regions. The topological Mott insulator with gapped  or gapless spin excitations can arise from the interplay of the Rashba SOC and correlations.    
\begin{description}
\item[Key words]Kane-Mele model; Rashba SOC-driven transition; strong correlations; slave-rotor mean field method.
\item[PACS numbers]73.20.At, 71.27.+a, 71.30.+h, 71.10.Fd

\end{description}
\end{abstract}

\maketitle


\section{\label{sec1:level1}Introduction}

Over the past few decades, topological insulators have attracted a lot of attentions from the condensed matter community, owing to its topologically nontrivial insulating band \cite{2010Hasan,2011Qi}. From a mathematical perspective, the novel structure of bands stems from the non-zero (actually an integer) consequence of the integral of the Berry curvature over half the Brillouin zone (BZ) or the integral winding number of the mapping from the Brillouin torus onto the space of Bloch Hamiltonians (a 2-sphere) \cite{2010Xiao,2010Wang,2007Moore,2003Avron}. The non-trivial consequence of the insulating band implies that with the change of their Hamiltonian systems belong to the same topological class in which these systems are adiabatically interconnected as long as the gap of their bands keeps opened. In other words, when the direct energy gap of a system loses, a topologcial insulating class possessed by the system can turn into the another one and the topological transition occurs. A significant toy model of topological insulators is the famous Kane-Mele (KM) model on the honeycomb lattice proposed by Kane and Mele in 2005 \cite{2005aKane}, and soon after its nontrivial band structure was characterized by the $Z_{2}$ topological invariant \cite{2005bKane}. The nontrivial band structrue of the model arises from the {\it intrinsic} spin-orbit coupling (SOC) of next-nearest neighbor electrons which is added to the tight-binding (TB) Hamiltonian of electrons in the graphene sheet to get the KM model. Although the time-reversal symmetry of the model is guaranteed by the intrinsic SOC, for each spin the symmetry is actually broken, i.e. the KM model can be thought of as two copies of the Haldane model \cite{1988Haldane} with opposite Chern numbers. It is the time-reversal symmetry breaking of each spin sector that leads to a non-trivial consequence of the integral of the Berry curvature over the BZ for bands in the KM model. Although the experimental observation of the topologically non-trivial insulating band possessed by the KM model has not been achieved due to the tiny magnitude of the intrinsic spin-orbit gap of graphene \cite{2007Yao}, the one possessed by the Bernevig-Hughes-Zhang model has been experimentally realized in the CdTe/HgTe/CdTe quantum well \cite{2006Bernevig,2007Konig}. 

Besides the intrinsic SOC, there is a so-called {\it extrinsic} SOC in 2D mesoscopic systems, i.e. the Rashba SOC \cite{1960Rashba,1984Bychkov}. It can have different origins, among which is the presence of external electric fields \cite{1984Bychkov}, a substrat \cite{2008Varyhalov} or neutral impurities \cite{2009Castro Neto,2011Weeks}. The effects of the Rashba SOC on the band of electrons in the graphene sheet have been investigated in detail\cite{2009Zarea,2009Rashba}. The extrinsic SOC can't open a gap in the energy band and maintains Dirac nodes of the original TB model of electrons in the graphene sheet. However, the spin degeneracy of each band is lifted due to the breaking of mirror symmetry, and then the Rashba SOC can force the electron in the graphene sheet to possess two zero-gap bands and two gapped bands. Furthermore, it is clear that the electron spin polarizations of all bands are in $(k_x,k_y)$-plane of momentum and they are perpendicular (or not) to momentum ${\bm k}$ depend on the Rashba SOC is isotropic (or not). Thus, It is interesting to observe what happens when the Rashba SOC is introduced into the the KM model---the electron model in the graphene sheet with intrinsic SOC which always possesses gapped spin-degenerate bands. Besides the splitting of bands, the Rashba SOC can remove the direct band gap caused by the intrinsic SOC if its strength is large enough \cite{2013Shakouri}. This remarkable consequence of the competition between the intrinsic SOC and the Rashba SOC is the destruction of the topologically non-trivial insulating band \cite{2005bKane,2005Sheng,2006Sheng}, i.e. the Rashba SOC can drive the KM model into a topologically trivial state from the $Z_{2}$ topological insulator. However, we suppose that the competition between the two SOCs in the KM model has not been explored fully. For example, the deformation of the band of the KM model caused by the Rashba SOC can lead to the shift of the touching point of the valence band and conduction band where the direct band gap closes and then the topological non-trivial transition occurs. Furthermore, the splitting of this touching point due to the crossing of the valence band and the conduction band have not been investigated yet. In this work, the effects of Rashba SOC on the band structure of KM model which have not been exposed entirely will be discussed in detail and the Rashba SOC-driven topologically trivial or non-trivial phase transition will be a main focus in the first part of our study.

For the topologically non-trivial insulating band, a natural and important question is to what extent  is its structure stable with respect to electron correlations and then what kinds of novel states can arise from the interplay between topology and electron correlations? Effects of electron correlations on topologically non-trivial insulating bands had already been investigated by several authors in the early years of topological insulators \cite{2005aKane,1984Niu,1985Niu}. The conclusion is that the topologically non-trivial insulating band is stable against the weaker electron correlation or disorder as long as they keep the gap opened. The two earlier investigations of the stronger electron correlation in topologically non-trivial insulating band were, in our opinion, given by Young \textit{et al.} \cite{2008Young} using the slave-rotor mean field theory and Cai \textit{et al.} \cite{2008Cai} using the Hartree-Fock mean field method. Since then, there have been a great deal of discussions on the effects of strong correlations on topological bands \cite{2013Hohenadler,2014Witczak-Krempa,2018Rachel}. Let us focus on the aspect of effects of strong correlations on the KM model on the honeycomb lattice. It has been investigated by various analytical or numerical methods, e.g. slave-particle/spin mean field theories \cite{2008Young,2010Rachel,2012Ruegg}, Schwinger boson/fermion approaches\cite{2012Vaezi}, the cellular dynamical mean field theory (DMFT) \cite{2012Wu}, the variational cluster approach (VCA) \cite{2011Yu} and the quantum Monte Carlo (QMC) simulation \cite{2011Hohenadler,2011Zheng,2012Hohenadler}. In general, the topologically non-trivial insulating band of the KM model is stable against the weaker correlations and the magnetic insulating phase which may destroy the topological structure of bands can emerge when the correlations become sufficiently strong. In the case of intermediate correlations, the topologically non-trivial structure of bands is maintained and various exotic states which stem from the interplay of topology and correlations, e.g. the topological Mott insulator (TMI), the quantum spin liquid (QSL), and the quantum spin Hall (QSH) state coupled to a dynamical Z$_{2}$ gauge field (QSH$^{*}$), can emerge. Some of the phase transitions in the correlated KM model have been investigated by Hohenadler \textit{et al.} \cite{2012Hohenadler} using the QMC simulation and Griset and Xu \cite{2012Griset} from the viewpoint of field theory. Furthermore, Bercx \textit{et al.} \cite{2014Bercx} have investigated effects of strong correlations on the KM model on the honeycomb lattice with a magnetic flux of $\pm \pi$ through each hexagon. They found that the antiferromagnetic order develops above a critical value of the correlation, which similar to the case of the ordinary correlated KM model, and there is a correlation-induced gap in the edge states as a result of umklapp scattering at half-filling.

Let's return to Rashba SOC in the case of electron correlations. At present, the investigations focus mainly on effects of the Rashba SOC on the edge states of topological insulators with correlation, e.g on the electron backscattering \cite{2010Strom,2012Schmidt,2012Budich}, and spin correlations and spectral gap \cite{2014Hohenadler} in correlated helical edge states. For the bulk of topological insulators, Laubach \textit{et al.} \cite{2014Laubach} reported, applying the variational cluster approach, a new topological-semiconductor phase which stems from the Rashba SOC in the Kane-Mele-Hubbard (KMH) model, and Mishra \textit{et al.} \cite{2018Mishra} sketched out the effects of the Rashba SOC on the phase diagram of interacting KM model at quarter filling. In the first part of our study here, the competition between the intrinsic SOC and the Rashba SOC has been investigated. We want to know more effects of electron correlations on the bulk of topological insulators with the Rashba SOC, especially on the Rashba SOC-driven topologically trivial or non-trivial phase transition. It is well known that the strong correlation can lead to a spin-charge separation where the charge degree of freedom is uncondensed and the spin degree of freedom may form a spin liquid state. In this work, focussing on the correlated KM model with the Rashba SOC, we will consider both of the spin-charge separation caused by the strong correlation and the topologically trivial or non-trivial phase transition driven by the Rashba SOC, and then investigate the influences of electron correlations on the Rashba SOC-driven phase transition in the condensated and Mott region of charge degree of freedom respectively. 

Our paper is organized as follows. In sec. \uppercase\expandafter{\romannumeral2}, we revisit the KM model with the Rashba SOC. The competition between the two SOCs is investigated in detail and the Rashba SOC-driven topologically trivial and non-trivial phase transition will be obtained from the energy spectrum. The value of the Rashba SOC at which the phase transition occurs is critical to the following discussion about the effects of correlation. In sec. \uppercase\expandafter{\romannumeral3}, the Hubbard interaction is introduced into the model and the metal-insulator transition of the charge degree of freedom is obtained by slave-rotor mean field method. In this section, we investigate in detail the phase transition driven by the Rashba SOC and Hubbard interaction. Finally, we conclude in sec. \uppercase\expandafter{\romannumeral4}.

\section{\label{sec2}The KM model with the Rashba SOC}
\subsection{\label{sec2-1}The model}

\begin{figure}[ht]
\includegraphics[width=8.5cm,height=4.3cm]{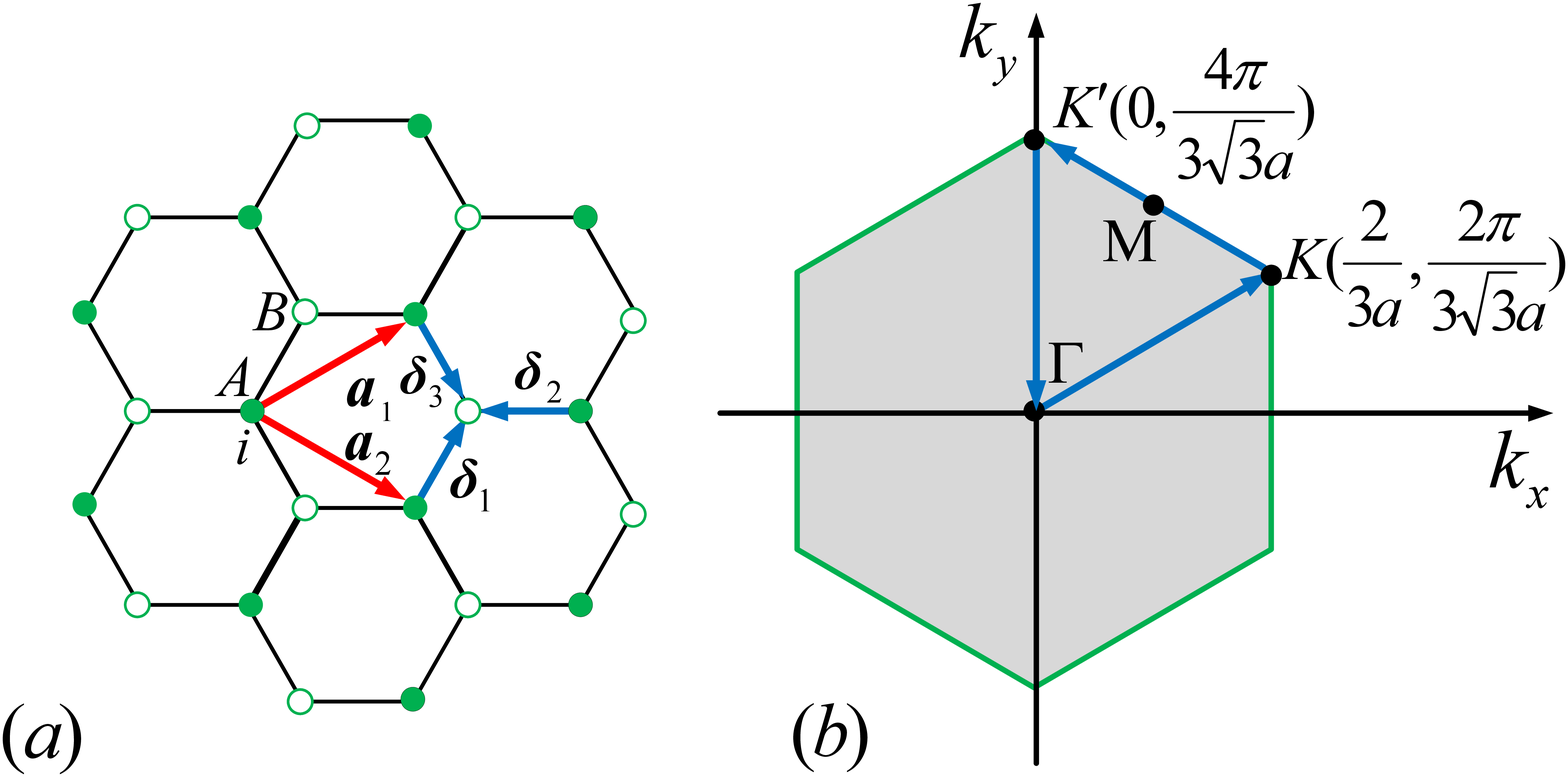}
\caption{\label{fig1}(Color online)(a) The honeycomb lattice. Red solid arrows represent the lattice vectors ${\bm a}_{1}=(3a/2, \sqrt{3}a/2)$ and ${\bm a}_{2}=(3a/2, -\sqrt{3}a/2)$. Blue solid arrows represent the NN bonds in three directions: ${\bm \delta}_{1}=(a/2, \sqrt{3}a/2)$, ${\bm \delta}_{2}=(a/2, -\sqrt{3}a/2)$ and ${\bm \delta}_{3}=(-a, 0)$. (b) The BZ of the model.}
\end{figure}

The KM model with the Rashba SOC (R-KM model) on the honeycomb lattice is
\begin{eqnarray}
H_{0}&=&-t\sum_{<ij>}\sum_{\sigma}\hat{c}^{\dagger}_{i\sigma}\hat{c}_{j\sigma}+\mathrm{i}\lambda\sum_{\ll ij\gg}\sum_{\sigma\sigma^{\prime}}\nu_{ij}\hat{c}_{i\sigma}^{\dagger}\sigma_{\sigma\sigma^{\prime}}^{z}\hat{c}_{j\sigma^{\prime}}\nonumber\\
&&+\mathrm{i}\alpha\sum_{<ij>}\sum_{\sigma\sigma^{\prime}}\hat{c}^{\dagger}_{i\sigma}(\bm{\sigma}_{\sigma\sigma^{\prime}}\times\bm{d}_{ij})^{z}\hat{c}_{i\sigma}.
\label{eq1}
\end{eqnarray}
Here the first term is the nearest neighbor (NN) electrons hopping term with the hopping strength $t$. The second term represents the intrinsic SOC between next-nearest neighbor (NNN) electrons with the coupling strength $\lambda$. $\sigma_{\sigma\sigma^{'}}^{z}$ is the $z$ component of Pauli matrices and the parameter $\nu_{ij}=-1$ if the orientation of the NNN sites $i$, $j$ is right turn while $\nu_{ij}=+1$ if left turn. The third term is the Rashba SOC term of NN electrons with coupling strength $\alpha$ and $\bm{d}_{ij}$ is the connected vector from site $i$ to site $j$. Lattice vectors of the honeycomb lattice are ${\bm a}_{1}=(3a/2, \sqrt{3}a/2)$ and ${\bm a}_{2}=(3a/2, -\sqrt{3}a/2)$, as shown in Fig.\ref{fig1}. In the concrete calculation, we set the lattice constant $a=1$ and the strength of NN hopping $t=1$.

In momentum space, the Hamiltonian of the KM model with the Rashba SOC can be obtained as 
\begin{eqnarray}
H_{0}=\sum_{\bm k}\Psi_{\bm k}^{\dagger}{\cal H}_{0\bm k}\Psi_{\bm k}.
\label{eq2}
\end{eqnarray}
Here $\Psi_{\bm k}=(\hat{c}_{{\bm k}\uparrow}^A, \hat{c}_{{\bm k}\uparrow}^B, \hat{c}_{{\bm k}\downarrow}^A, \hat{c}_{{\bm k}\downarrow}^B)^T $ is the matrix of electron operators in the momentum-spin space and the Bloch Hamiltonian ${\cal H}_{0\bm k}$ is
\begin{eqnarray}
\begin{pmatrix}
\lambda \gamma&-tg&0&\!\alpha(w_{x}\!-\!\mathrm{i}w_{y})\\
-tg^{*}&-\lambda \gamma&\!\alpha(w_{x}^{*}\!-\!\mathrm{i}w_{y}^{*})&0\\
0&\!\alpha(w_{x}\!+\!\mathrm{i}w_{y})&-\lambda \gamma&-tg\\
\!\alpha(w_{x}^{*}\!+\!\mathrm{i}w_{y}^{*})&0&-tg^{*}&\lambda \gamma
\end{pmatrix},
\label{eq3}
\end{eqnarray}
where A, B represent the sublattice of the honeycomb lattice as shown in Fig.~\ref{fig1},  $g=\sum_{i}^{3}e^{\mathrm{i}{\bm k}\cdot\bm \delta_{i}}$, $w_{x}=\mathrm{i}\sqrt{3}a(e^{\mathrm{i}\bm{k}\cdot\bm{\delta}_{1}}-e^{\mathrm{i}\bm{k}\cdot\bm{\delta}_{2}})/2$, $w_{y}=\mathrm{i}a(-e^{\mathrm{i}\bm{k}\cdot\bm{\delta}_{1}}-e^{\mathrm{i}\bm{k}\cdot\bm{\delta}_{2}}+2e^{\mathrm{i}\bm{k}\cdot\bm{\delta}_{3}})/2$ and  $\gamma=2[-\sin(\sqrt{3}ak_{y})+2\cos(3ak_{x}/2)\sin(\sqrt{3}ak_{y}/2)]$.

\subsection{\label{sec2-2}The competition between the intrinsic and Rashba SOC and Rashba SOC-driven transitions }

It is the non-zero elements at counter-diagonal of Hamiltonian matrix ${\cal H}_{0\bm k}$ that mix the up-pin and down-spin and violate mirror symmetry, and then lift the spin degeneracy of bands and break the particle-hole symmetry of bands. Although these terms prevent us from diagonalizing the Hamiltonian by hand easily, the band energy and eigenstates of bands can be obtained numerically or analytically by computer.    

For $\lambda=0$ and $\alpha=0$, the energy band is just the one of electrons in the graphene sheet with the spin, which is gapless and has Dirac nodes at $K(2\pi/3a,2\pi/3\sqrt{3}a)$ and $K^{\prime}(0,4\pi/3\sqrt{3}a)$ in the BZ. For $\lambda=0$ and $\alpha\not=0$, the Rashba SOC lifts the spin degeneracy of bands and keeps the gapless of the energy spectrum. For $\lambda\not=0$ and $\alpha=0$, bands of the KM model have the spin degeneracy and the intrinsic SOC always opens energy gaps at $K$ and $K^{\prime}$. The previous results are well known in the graphene research community. The case of $\lambda\not=0$ and $\alpha\not=0$ where the Rashba SOC competes with the intrinsic SOC will be investigated as follows. In our discussions, a ratio $\chi$ is defined as $\alpha/\lambda$.

\begin{figure}[ht]
\includegraphics[width=4.25cm,height=3.6cm]{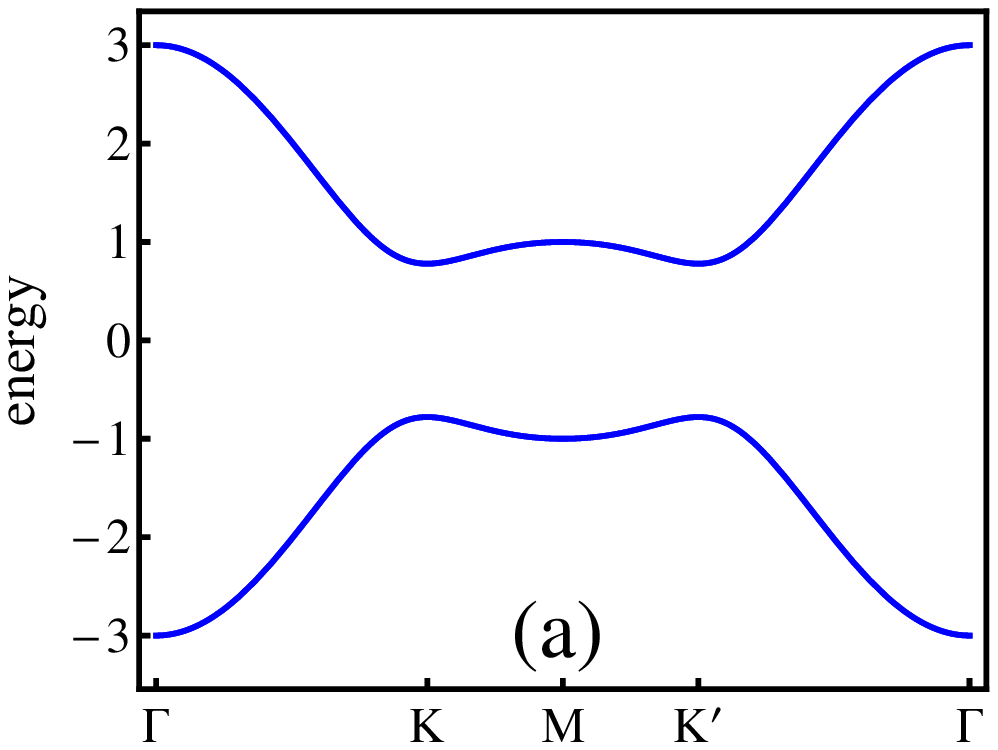}
\includegraphics[width=4.25cm,height=3.6cm]{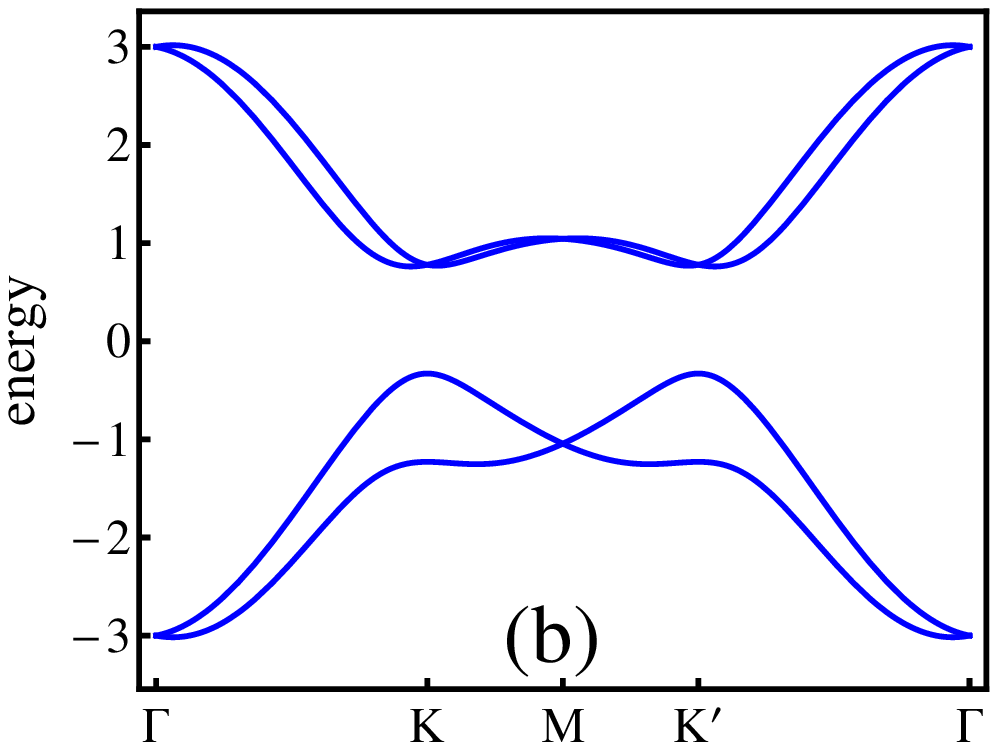}
\includegraphics[width=4.25cm,height=3.6cm]{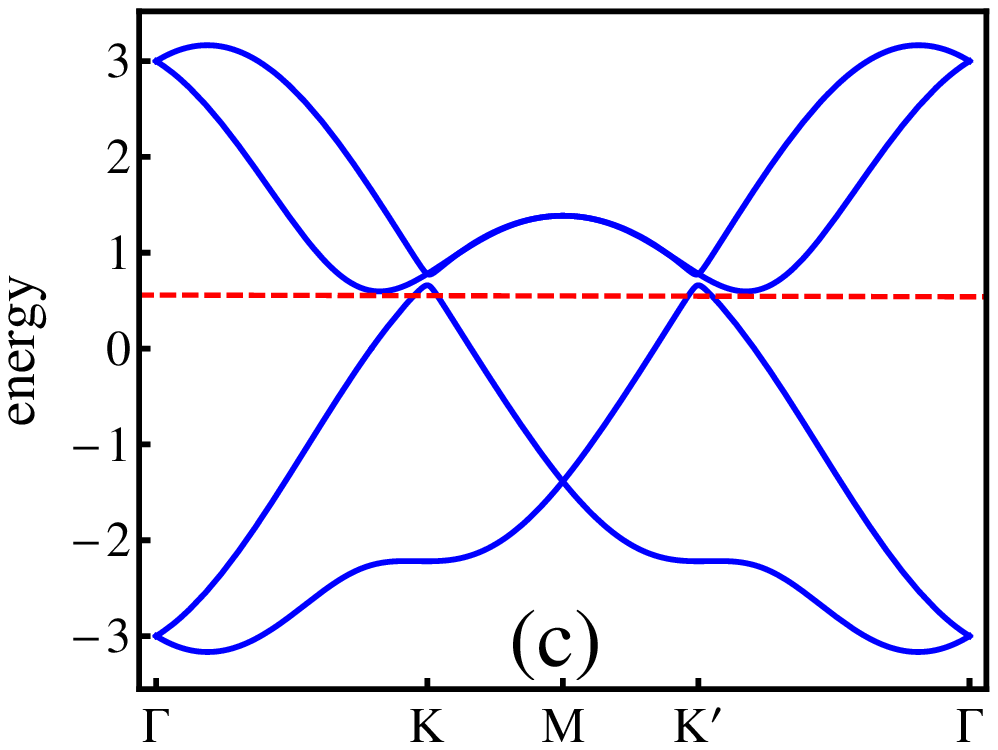}
\includegraphics[width=4.25cm,height=3.6cm]{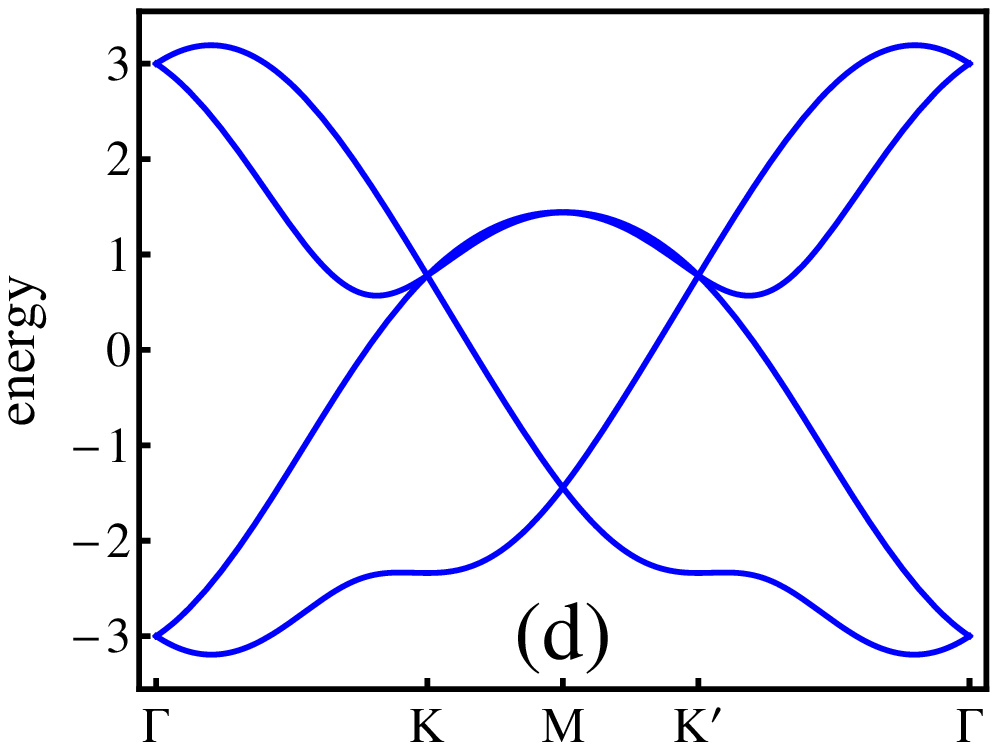}
\caption{\label{fig2}(Color online)Effects of Rashba SOC on energy band structure of the KM model at the strength of intrinsic SOC $\lambda=0.1$. The path in the BZ is taken as shown in the Fig.\ref{fig1}(b). (a) $\chi=0$; (b) $\chi<\chi_{ntopo}\approx3.273$; (c) $\chi>\chi_{ntopo}$; (d) $\chi=\chi_{topo}=\chi_{0}$ ($\chi_{0}=2\sqrt{3}$).}
\end{figure}

Firstly, let us focus on the case of the small intrinsic SOC. When the Rashba SOC increases gradually at $\lambda\not=0$, besides the lifting of the spin degeneracy of bands, the band gap decreases and the touching of the valence band and the conduction band occurs eventually. It is clear that, at the smaller intrinsic SOC, the touching points are located at $K$ and $K^{\prime}$ in the BZ due to the C$_{3}$ symmetry and Dirac cones form at these touching points. The famous result \cite{2005bKane} of the critical value of Rashba SOC at which the band touching or the topologically non-trivial transition from the $Z_{2}$ topological insulator to a trivial matter occurs is $\alpha=\chi_{topo}\cdot\lambda$ , where $\chi_{topo}=\chi_{0}\overset{\text{def}}{=}2\sqrt{3}$. We find that, before $\chi$ reaches the critical value $\chi_{0}$, the indirect band gap had already closed at some critical values of the Rashba SOC $\alpha=\chi_{ntopo}\cdot\lambda$. When the $\chi$ exceeds the critical value $\chi_{ntopo}$, the indirect overlap of bands occurs as shown in Fig.\ref{fig2}. 
It leads to a density of holes in the ``valence band" and electrons in the ``conduction band", and then provides charge carriers for electrical condutivity at low temperatures. In the case, the phase is a semi-metal (SM) state \cite{1997Gebhard}. On the other hand, the $Z_{2}$ topological invariant is still well-defined because of the non-contact of the bands. Thus, the closing of the indirect gap indicates a topologically trivial metal-band insulator transition \cite{1997Gebhard} driven by the Rashba SOC and we call the semi-metal a ``topological semi-metal"(TSM).    

\begin{figure}[ht]
\includegraphics[width=4.25cm,height=4.25cm]{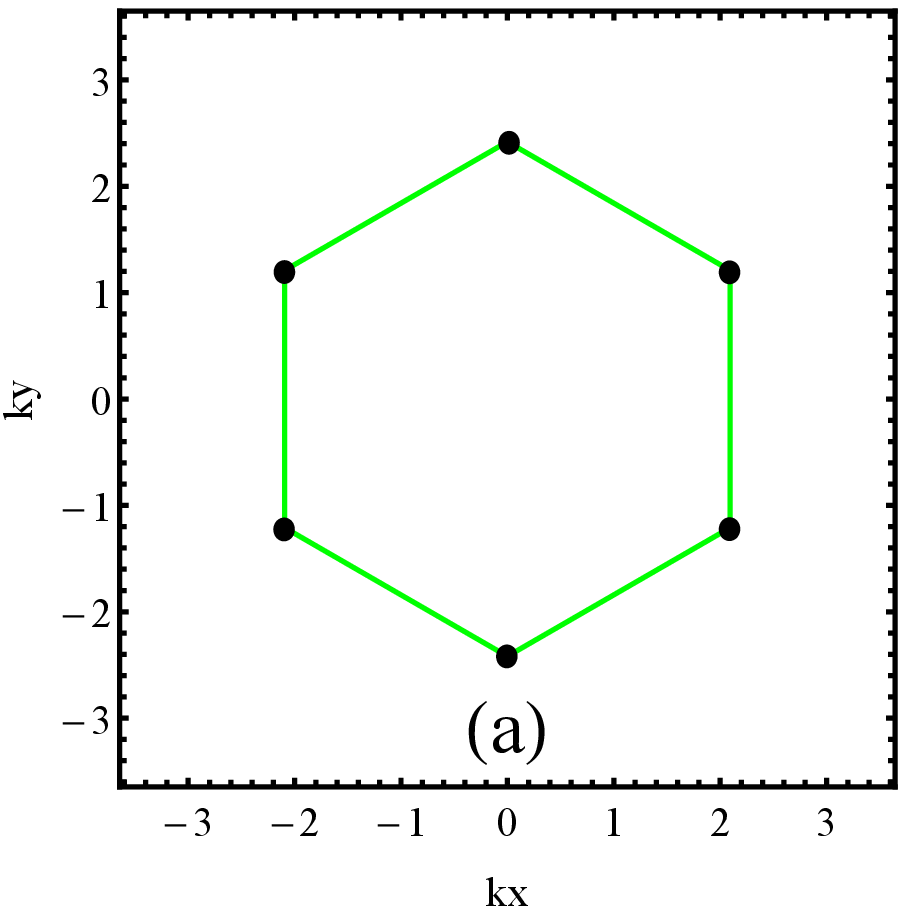}
\includegraphics[width=4.25cm,height=4.25cm]{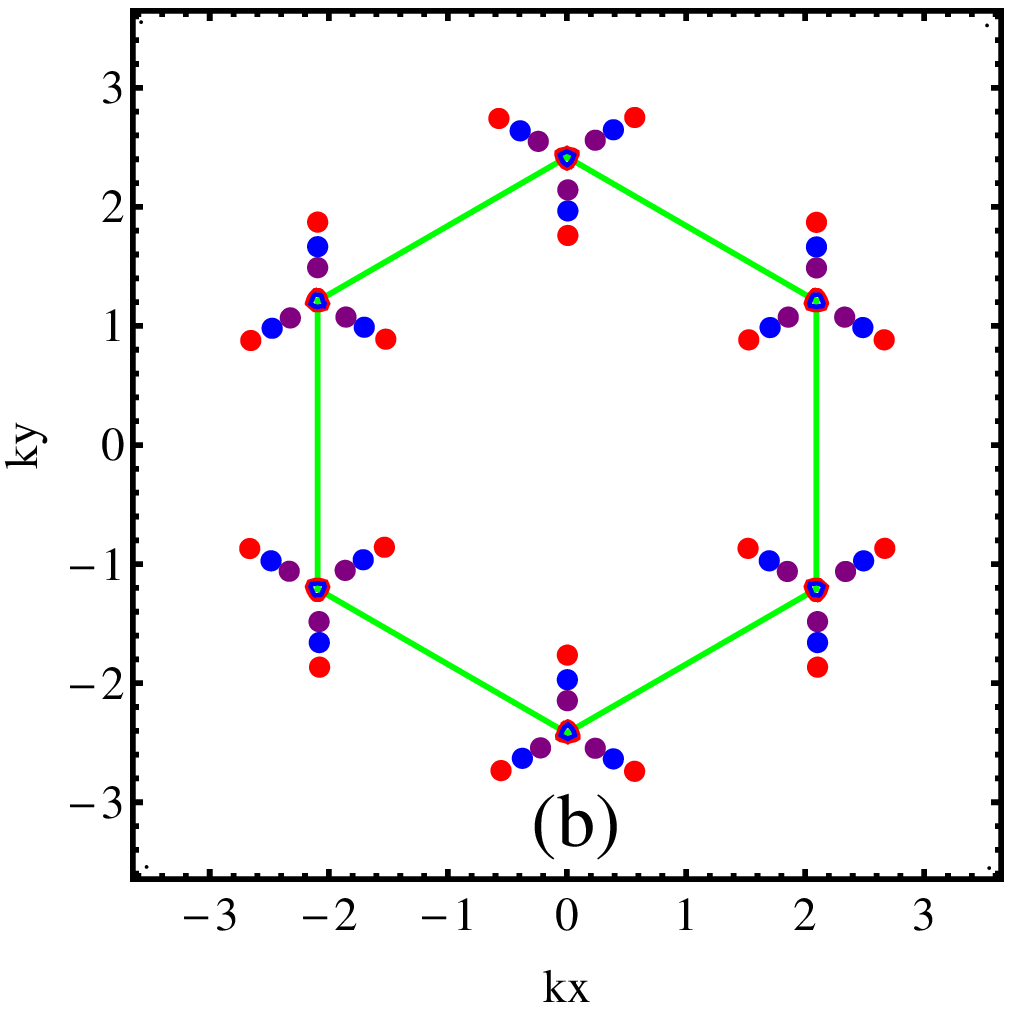}
\caption{\label{fig3}(Color online)Locations of touching points of the valence band and the conduction band in BZ at $\lambda=0.1$. (a) The case of $\chi=\chi_{topo}=\chi_{0}$ ($\chi_{0}=2\sqrt{3}$). The black points represent the locations of touching points, i.e. $K$ and $K^{\prime}$. (b) The case of $\chi>\chi_{0}$. The purple, blue and red points represent the locations of touching points when $\chi=5, 6$ and $8$ respectively. Points $K$ and $K^{\prime}$ are both the locations of touching points for all of the three  $\chi$s.}
\end{figure}

To further observe the competition between Rashba and intrinsic SOC, it is valuable to let the larger Rashba SOC to dominate. We find that the gapless of bands maintains and there are more touching points of the valence band and the conduction band in the BZ, besides the ones at $K$ and $K^{\prime}$ where the topologically non-trivial transition occurs. The splitting of original touching pionts at $K$ and $K^{\prime}$ is shown in Fig.\ref{fig3}. Around each point $K$ or $K^{\prime}$, there are three additional touching points which have $2\pi/3$ rotational symmetry possessed by the graphene lattice of the KM model with the Rashba SOC. Furthermore, the locations of these additional points will deviate from the $K$ or $K^{\prime}$ with the increasing Rashba SOC. The splitting of touching point is the consequence of crossings of valence and conduction bands when the Rashba SOC dominates in the competition. The splitting of the touching point caused by crossings of the two bands is actually an extension of effects of the Rashba SOC on the Dirac point of graphene energy spectrum as disscussed by Zarea and Sandler \cite{2009Zarea}.

So far, we have a basic scenario about the competition between the intrinsic SOC and the Rashba SOC by increasing gradually the Rashba SOC at some small intrinsic SOCs (e.g. $\lambda=0.1$). When the intrinsic SOC dominates, the KM model with Rashba SOCs is a $Z_2$ topological insulator. At a critical Rashba SOC, i.e. $\alpha=\chi_{ntopo}\cdot\lambda$ ($\chi_{ntopo}\approx3.273$ for $\lambda=0.1$) the indirect gap of the valence band and conduction band closes and a Rashba SOC-driven topologically trivial transition from the $Z_2$ topological insulator to the TSM occurs. The topologically non-trivial phase keeps until the band touching occurs at points $K$ and $K^\prime$. The Rashba SOC-driven topologically non-trivial transition (or band touching) occurs when the Rashba SOC reaches the second critical value $\alpha=\chi_{topo}\cdot\lambda$, where $\chi_{topo}=\chi_{0}=2\sqrt{3}$. Beyond the transition, the system stays in a topologically trivial metal phase because of the gapless of energy band and has more band touching points which split from the ones at $K$ and $K^{\prime}$ due to the dominance of the Rashba SOC. 

\begin{figure}[ht]
\includegraphics[width=4.25cm,height=4.25cm]{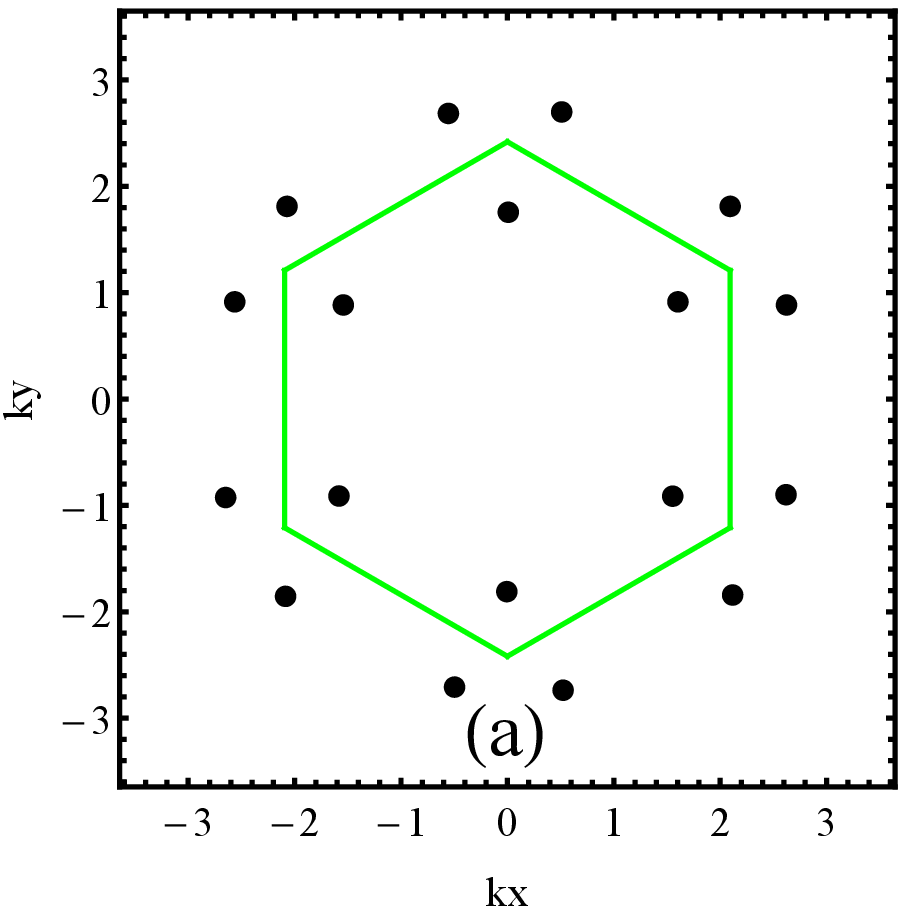}
\includegraphics[width=4.25cm,height=4.25cm]{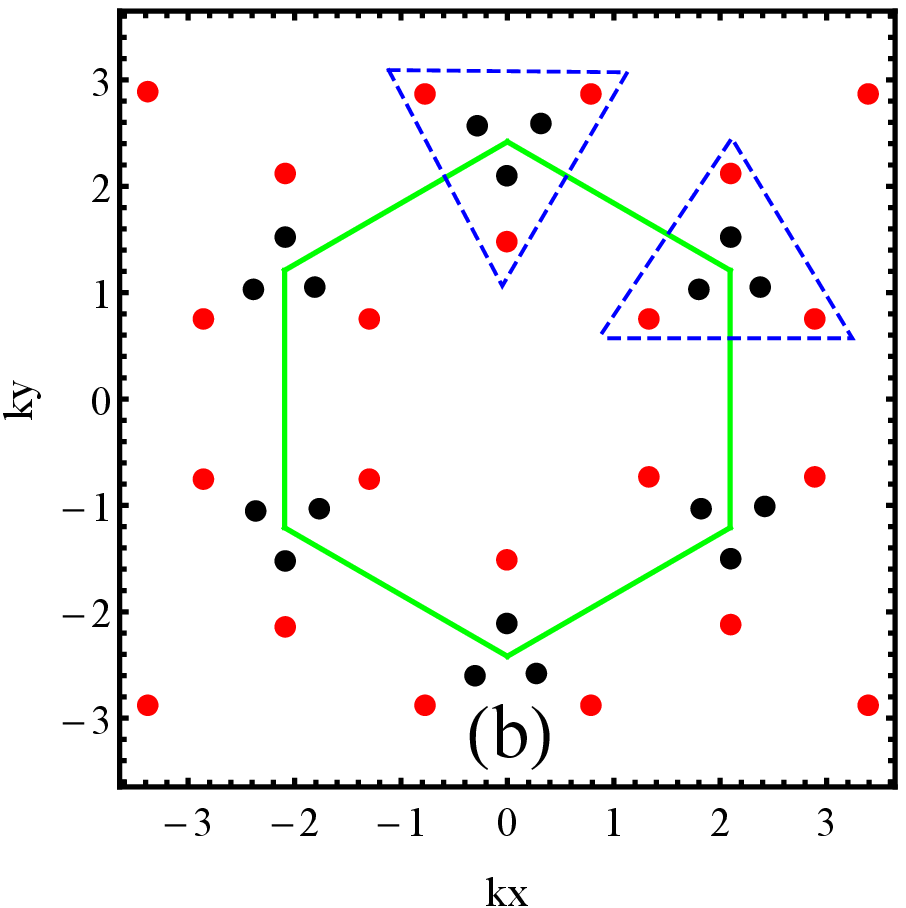}
\includegraphics[width=4.25cm,height=4.25cm]{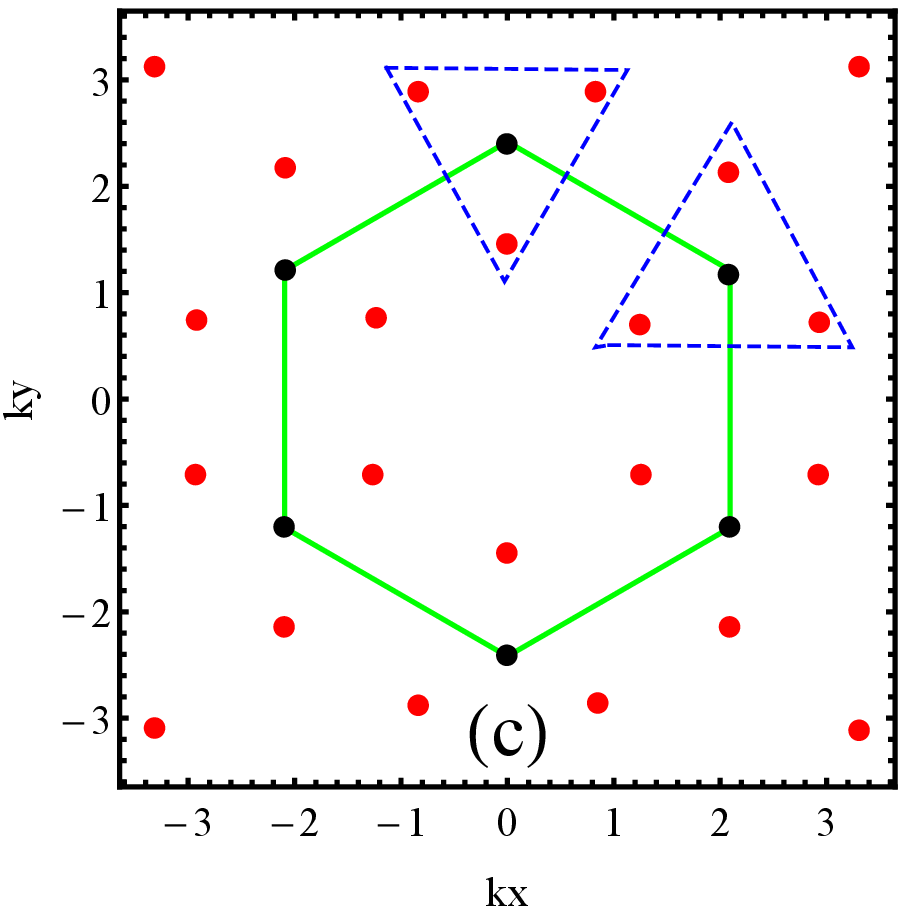}
\includegraphics[width=4.25cm,height=4.25cm]{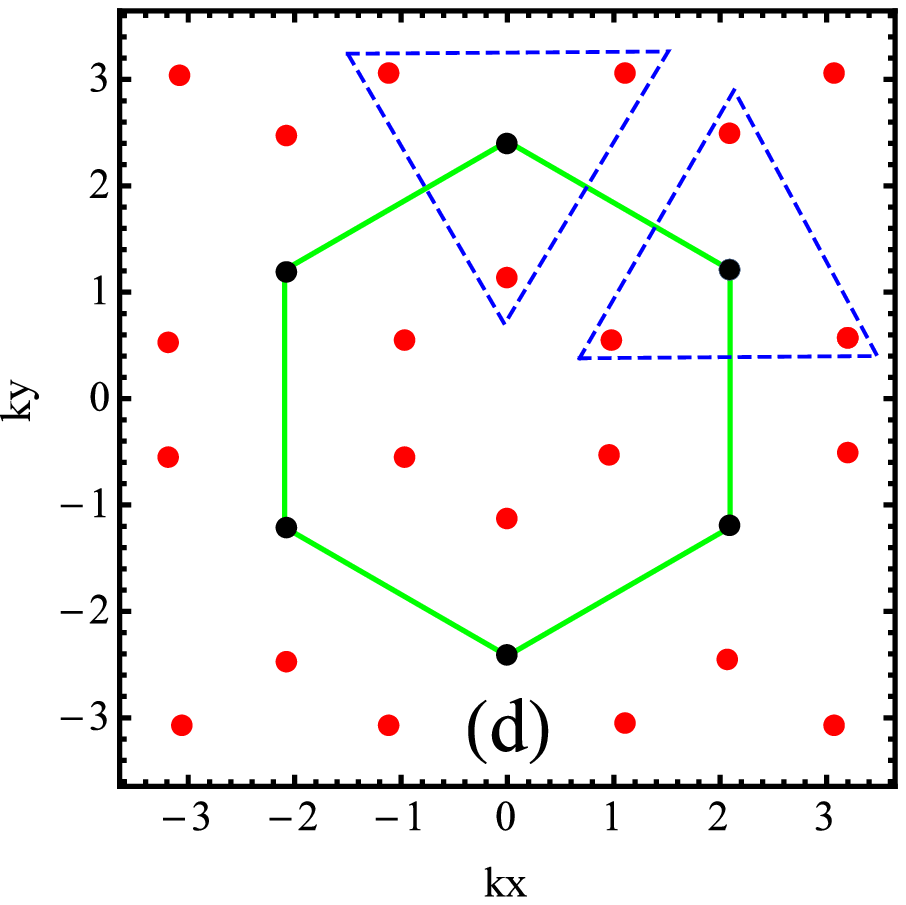}
\caption{\label{fig4}(Color online)Locations of touching points of the valence band and the conduction band in BZ at $\lambda=0.4$ in the case of $\chi\ge\chi_{topo}\approx2.903$. (a) $\chi=\chi_{topo}$, (b) $\chi_{topo}<\chi<\chi_{0}$, (c) $\chi=\chi_{0}$ and (d) $\chi>\chi_{0}$. The blue dotted triangles are used to mark out the locations around the $K$ and $K^{\prime}$.} 
\end{figure}

However, this is not the full story. The missing piece can be retrieved from the investigation of band touching of the valence and conduction bands at large intrinsic SOCs. It is also a fact that the band touching at $K$ and $K^{\prime}$ always occurs when the Rashba SOC $\alpha=\chi_{0}\lambda$ ($\chi_{0}=2\sqrt{3}$). Therefore, the larger the intrinsic SOC is, the larger Rashba SOC is needed to develop this type of band touching. On the other hand, the large Rashba SOC can greatly ``enhance" the crossing of bands which causes the additional band touching. Thus it is possible that the additional band touching which differs from the one at $K$ or $K^{\prime}$ can occur before the Rashba SOC reaches the critical value of $\chi_{0}\lambda$. The procees in the case of $\lambda=0.4$ is shown in the Fig.\ref{fig4}. 
At the critical Rashba SOC $\alpha=\chi_{topo}\cdot\lambda$ ($\chi_{topo}\approx2.903<\chi_{0}$), the band touching occurs at the some points who differ from the $K$ and $K^{\prime}$. These locations of band touching points in BZ also possess the $2\pi/3$ rotational symmetry as shown in Fig.\ref{fig4}(a). Here we assign the critical $\chi$ as $\chi_{topo}$, because the band touching accompanies a topologically non-trivial transition due to the closing of the direct band gap. Beyond this critical Rashba SOC, the model stays in the topologically trivial metal state and the effect of the Rashba SOC is just to split and shift the touching points (see below). When the Rashba SOC increases further ($\chi>\chi_{topo}$), touching points are split because of the band crossing as shown in Fig.\ref{fig4}(b). The inside points move towards the $K$ or $K^{\prime}$ with the increasing Rashba SOC. Eventually, a touching point forms at $K$ or $K^{\prime}$ when $\chi=\chi_{0}$ and there are three satellite touching points around the single point, which is similar to the case of small intrinsic SOCs, e.g. $\lambda=0.1$. The case of $\chi=\chi_{0}$ is shown in Fig.\ref{fig4}(c). The further increasing Rashba SOC enforces satellite touching points to move away from points $K$ or $K^{\prime}$ and maintains the single touching point at the $K$ or $K^{\prime}$ as shown in Fig.\ref{fig4}(d). 

For the small Rashba SOC ($\chi<\chi_{topo}$), the intrinsic SOC dominates and the model possesses topologically non-trivial states. There is also a Rashba SOC-driven topologically trivial transition from the $Z_2$ topological insulator to the TSM at a critical value $\chi=\chi_{ntopo}$ ($\chi_{ntopo}\approx2.384$ in the case of $\lambda=0.4$) and the Rashba SOC also causes the indirect overlap between the valence band and the conduction band when $\chi>\chi_{ntopo}$. The band structures are very similar to the ones at $\lambda=0.1$ as shown in Fig.\ref{fig2}(a)--(c) and we do not draw them here.

\begin{figure}[h]
\includegraphics[width=8.5cm,height=5.5cm]{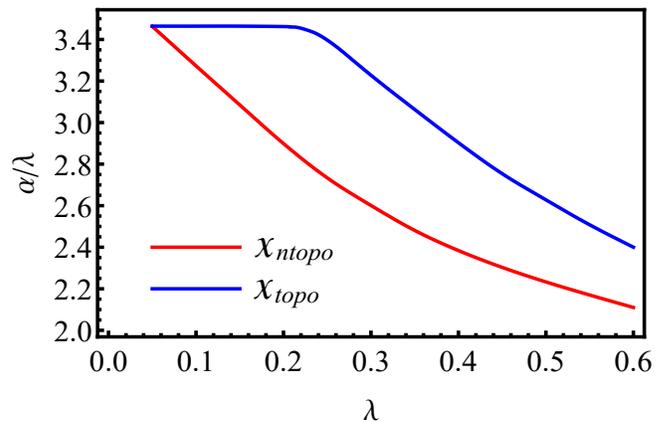}
\caption{\label{fig5}(Color online)Critical value of ratio $\chi=\alpha/\lambda$ at which the topologically trivial or non-trivial transition occurs. For small enough intrinsic SOC, the $\chi_{topo}=2\sqrt{3}$ as the classical result.} 
\end{figure}

\begin{figure}[h]
\includegraphics[width=8.5cm,height=2cm]{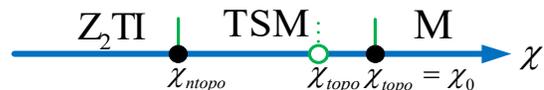}
\caption{\label{fig6}(Color online)The phase diagram of the R-KM model. $Z_{2}$TI: $Z_{2}$ topological insulator, TSM: topological semi-metal and M: metal. The critical value $\chi_{topo}$ =$\chi_{0}=2\sqrt{3}$ for the smaller intrinsic SOC and $\chi_{topo}<\chi_{0}$ for the larger one. The variation of $\chi_{ntopo}$ or $\chi_{topo}$ with the increasing intrinsic SOC was shown in Fig.\ref{fig5}.} 
\end{figure}

The above investigations on the competition between the intrinsic SOC and the Rashba SOC can be carried out in principle for all of the intrinsic SOC. Rashba SOC-driven phase transitions can also be obtained from the closing of the indirect band gap or/and the direct band gap (corresponding to the band touching). The critical values of ratio $\chi$ at which the topologically trivial or non-trivial transition occurs for various intrinsic SOCs are shown in Fig.\ref{fig5}. Here, we also draw the phase diagram of the R-KM model as shown in Fig.\ref{fig6}. 

\section{\label{sec3}The R-KM model with strong correlations}

In the section, the strong correlation represented by the on-site Hubbard interaction is introduced into the R-KM model (called R-KMH model). Strongly correlated systems can display the spin-charge separation. It postulates that electrons in these systems can be viewed as composites of chargons and spinons. The slave-rotor representation of the physical electron operators can treat economically the spin-charge separation and describe appropriately the Mott transition of the charge degree of freedom in the region of intermediate correlations\cite{2002Florens,2003Florens,2004Florens}. It is well known that the correlation can renormalize model parameters of a system and then change the physical quantities of the system. In the slave-rotor mean field method the benefit is that the spin degree of freedom (i.e. the spinon) inherits the Hamiltonian of physical electrons, but with the renormalized model parameters, e.g. the renormalized intrinsic or Rashba SOC in our studied model here. In the preceding section we obtained the Rashba SOC-driven topologically trivial and non-trivial transition. Therefore, it is interesting to observe what happens to the Rashba SOC-driven transition when the R-KM model is renormalized by the correlation. Here we capture the Mott transition of the charge degree of freedom at the intermediate Hubbard interaction using the slave-rotor mean field method. The boundary of the Mott transition is obtained numerically from the mean field self-consistency equations in this section. Furthermore, the slave-rotor mean field method will be applied to obtain Rashba SOC-driven topologically trivial and non-trivial transitions in both cases of condensed and uncondensed charges. We will compare these results with the ones in the case of non-interacting limit. The influences of correlations on the Rashba SOC-driven topologically trivial and non-trivial transitions in regions of charge condensate (i.e. for physical electron) and charge uncondensate (i.e. for spinon only) are discussed in detail here.

\subsection{\label{sec3-1}The slave-rotor mean field method for correlated R-KM model}
\subsubsection{\label{sec3-1-1}Slave-rotor representation for the correlated model}

When the on-site Hubbard term is introduced, the Hamiltonian of the R-KMH model reads
\begin{equation}
H=H_{0}+\frac{U}{2}\sum_{i}\Big(\sum_{\sigma}n_{i\sigma}-1\Big)^{2}.
\label{eq4}
\end{equation}
Here $H_{0}$ is given by Eq.(\ref{eq1}) and $n_{i\sigma}=\hat{c}_{i\sigma}^{\dagger}\hat{c}_{i\sigma}$ is the number operator of electrons with spin-$\sigma$.

The slave-rotor representation\cite{2002Florens,2003Florens,2004Florens} decomposes the physical electron annihilation operator as
\begin{eqnarray}
\hat{c}_{i\sigma}=e^{\mathrm{i}\theta_{i}}\hat{f}_{i\sigma}.
\label{eq5}
\end{eqnarray}
Here $e^{\mathrm{i}\theta_{i}}$ is the U(1) rotor operator that describes the charge degree of freedom and  $\hat{f}_{i\sigma}$ is the spinon operator that describes the spin degree of freedom of the electron. There is a constraint that recover the Hilbert space of the electron
\begin{eqnarray}
\sum_{\sigma}\hat{f}_{i\sigma}^{\dagger}\hat{f}_{i\sigma}+ \hat{L}_{i}=1.
\label{eq6}
\end{eqnarray}
Where the canonical angular momentum $\hat{L}_{i}=\mathrm{i}\partial_{\theta_{i}}$ associated with the angular $\theta_{i}$ is introduced. 

The model Hamiltonian can be written in the rotor and spinon operators as
\begin{eqnarray}
H&=&-t\sum_{<ij>\sigma}e^{-\mathrm{i}\theta_{ij}}\hat{f}^{\dagger}_{i\sigma}\hat{f}_{j\sigma}\nonumber\\
&&+\mathrm{i}\lambda\sum_{\ll ij\gg}\sum_{\sigma\sigma^{\prime}}\nu_{ij}e^{-\mathrm{i}\theta_{ij}}\hat{f}_{i\sigma}^{\dagger}\sigma_{\sigma\sigma^{\prime}}^{z}\hat{f}_{j\sigma^{\prime}}\nonumber\\
&&+\mathrm{i}\alpha\sum_{<ij>}\sum_{\sigma\sigma^{\prime}}e^{-\mathrm{i}\theta_{ij}}\hat{f}_{i\sigma}^{\dagger}(\bm{\sigma}_{\sigma\sigma^{\prime}}\times\bm{d}_{ij})^{z}\hat{f}_{j\sigma^{\prime}}\nonumber\\
&&+\frac{U}{2}\sum_{i}\hat{L}_{i}^{2}-\mu\sum_{i\sigma}\hat{f}_{i\sigma}^{\dagger}\hat{f}_{j\sigma}.
\label{eq7}
\end{eqnarray}
Here, $\theta_{ij}=\theta_{i}-\theta_{j}$, and $\mu$ is the chemical potential. The partition function of the system is written as a path integral of $e^{-S_{E}}$ over fields $f$, $f^{*}$ and $\theta$, where
\begin{eqnarray}
S_{E}&=&\int_{0}^{\beta}d\tau\Big[\sum_{i}-\mathrm{i}L_{i}\partial_{\tau}\theta_{i}+\sum_{i\sigma}f^{*}_{i\sigma}\partial_{\tau}f_{i\sigma}+H\nonumber\\
&&+\sum_{i}h_{i}\Big(\sum_{\sigma}f^{*}_{i\sigma}f_{i\sigma}+L_{i}-1\Big)\Big]
\label{eq8}
\end{eqnarray}
is the action in \textit{imaginary time} ($\tau=\mathrm{i}t$). Here the last term is introduced into the action to satisfy the constraint of Eq.(\ref{eq6}). From the canonical equation of motion in imaginary time, i.e. $\mathrm{i}\partial_{\tau}\theta_{i}=\partial H/\partial L_{i}$, we can obtain the relation of $L$ and $\theta$ as $L_{i}=(\mathrm{i}/U)\partial_{\tau}\theta_{i}$. Therefore, considering the Hamiltonian Eq.(\ref{eq7}) the action of the R-KMH model becomes
\begin{eqnarray}
S_{E}&=&\int_{0}^{\beta}d\tau\Big[\sum_{i\sigma}f^{*}_{i\sigma}(\partial_{\tau}-\mu+h_{i})f_{i\sigma}+\sum_{i}(-h_{i}+\frac{h_{i}^{2}}{2U})\nonumber\\
&&+\frac{1}{2U}\sum_{i}(\partial_{\tau}\theta_{i}+\mathrm{i}h_{i})^{2}-t\sum_{<ij>\sigma}e^{-\mathrm{i}\theta_{ij}}f^{*}_{i\sigma}f_{j\sigma}\nonumber\\
&&+\mathrm{i}\lambda\sum_{\ll ij\gg}\sum_{\sigma\sigma^{\prime}}\nu_{ij}e^{-\mathrm{i}\theta_{ij}}f_{i\sigma}^{*}\sigma_{\sigma\sigma^{\prime}}^{z}f_{j\sigma^{\prime}}\nonumber\\
&&+\mathrm{i}\alpha\sum_{<ij>}\sum_{\sigma\sigma^{\prime}}e^{-\mathrm{i}\theta_{ij}}\hat{f}_{i\sigma}^{\dagger}(\bm{\sigma}_{\sigma\sigma^{\prime}}\times\bm{d}_{ij})^{z}\hat{f}_{j\sigma^{\prime}}\Big].
\label{eq9}
\end{eqnarray}
It is more convenient to introduce a new field $X_{i}=e^{\mathrm{i}\theta_{i}}$ to represent the charge degree of freedom. The new field satisfies the constraint $|X_{i}|^2=1$ due to its complex exponential form. Furthermore, to express the action in quadratic form of $X$-field and $f$ field five mean field parameters should be introduced:
\begin{eqnarray}
&&Q_{X}=\big<\sum_{\sigma}f_{i\sigma}f_{j\sigma}\big>_{<ij>},
\label{eq10}\\
&&Q_{f}=\big<e^{-\mathrm{i}\theta_{ij}}\big>_{<ij>}=\big<X_{i}^{*}X_{j}\big>_{<ij>},
\label{eq11}\\
&&Q_{X}^{\prime}=\big<\sum_{\sigma\sigma^{\prime}}\mathrm{i}\nu_{ij}f_{i\sigma}^{*}\sigma_{\sigma\sigma^{\prime}}^{z}f_{j\sigma^{\prime}}\big>_{\ll ij \gg},
\label{eq12}\\
&&Q_{f}^{\prime}=\big<e^{-\mathrm{i}\theta_{ij}}\big>_{\ll ij \gg}=\big<X_{i}^{*}X_{j}\big>_{\ll ij \gg},
\label{eq13}\\
&&Q_{X}^{\prime\prime}=\big<\sum_{\sigma\sigma^{\prime}}\mathrm{i}(f_{i\sigma}^{*}(\bm{\sigma}_{\sigma\sigma^{\prime}}\times\bm{d}_{ij})^{z}f_{j\sigma^{\prime}}\big>_{<ij>}.
\label{eq14}
\end{eqnarray}
Then, the action can be obtained as
\begin{eqnarray}
S_{E}&=&\int_{0}^{\beta}d\tau\Big[\frac{1}{2U}\sum_{i}\mathrm{i}\partial_{\tau}X_{i}^{*}(-\mathrm{i}\partial_{\tau})X_{i}+\sum_{i}\rho_{i}|X_{i}|^{2}+H^{X}\nonumber\\
&&+\sum_{i\sigma}f_{i\sigma}^{*}\partial_{\tau}f_{i\sigma}+H^{f}+\cdots\Big].
\label{eq15}
\end{eqnarray}
Here, the symbol $``\cdots"$ denotes constant terms of mean field decomposition and we have set $h_{i}\equiv h=-\mu=0$ for half-filling at the mean field level. $\rho_{i}$ is the Lagrange multiplier for constraint $|X_{i}|^2=1$ and $\rho_i\equiv\rho$ in the mean field treatment. In the above expression,
\begin{eqnarray}
H^{X}&=&-tQ_{X}\sum_{<ij>}X_{i}^{*}X_{j}+\lambda Q_{X}^{\prime}\sum_{\ll ij \gg}X_{i}^{*}X_{j}\nonumber\\
&&+\alpha Q_{X}^{\prime\prime}\sum_{<ij>}X_{i}^{*}X_{j}
\label{eq16}\end{eqnarray}
and 
\begin{eqnarray}
H^{f}&=&-tQ_{f}\sum_{<ij>\sigma}f_{i\sigma}^{*}f_{j\sigma}+\mathrm{i}\lambda Q_{f}^{\prime}\sum_{\ll ij \gg}\sum_{\sigma\sigma^{\prime}}\nu_{ij}f_{i\sigma}^{*}\sigma_{\sigma\sigma^{\prime}}^{z}f_{j\sigma^{\prime}}\nonumber\\
&&+\mathrm{i}\alpha Q_{f}\sum_{<ij>}\sum_{\sigma\sigma^{\prime}}f_{i\sigma}^{*}(\bm{\sigma}_{\sigma\sigma^{\prime}}\times\bm{d}_{ij})^{z}f_{j\sigma^{\prime}}.
\label{eq17}
\end{eqnarray}
The action of Eq.(\ref{eq15}) can be transformed into frequency-momentum space via Fourier transforms
\begin{eqnarray}
&&X_{i}(\tau)=\frac{1}{\sqrt{\beta N_{\Lambda}}}{\sum_{\bm{k},n}}'e^{\mathrm{i}(\bm{k}\cdot\bm{R}_{i}-v_{n}\tau)}X_{\bm{k}}(\mathrm{i}v_{n})+\sqrt{x_{0}}
\label{eq18},\\
&&f_{i\sigma}(\tau)=\frac{1}{\sqrt{\beta N_{\Lambda}}}{\sum_{\bm{k},n}}e^{\mathrm{i}(\bm{k}\cdot\bm{R}_{i}-\omega_{n}\tau)}f_{\bm{k}\sigma}(\mathrm{i}\omega_{n})
\label{eq19}.
\end{eqnarray}
Here $N_{\Lambda}$ denotes the number of unit cells and $x_{0}$ is the density of the condensate of charges. $v_{n}=2n\pi/\beta$ are the Matsubara frequencies for bosons and $\omega_{n}=(2n+1)\pi/\beta$ for fermions and the summation excludes the point $(\mathrm{i}v_{n}^0,\bm{k}^0)$ at which the condensate of charges occurs. Finally, we can write the action in matrix form as
\begin{eqnarray}
S_{E}&=&\sum_{\bm{k},n}\Psi_{\eta}^{X\dagger}\Big[\Big(\frac{v_{n}^{2}}{2U}+\rho\Big)\delta_{\eta\kappa}+{\cal{H}}_{\eta\kappa}^{X}\Big]\Psi_{\kappa}^{X}\nonumber\\
&&+\sum_{\bm{k},n}\Psi_{\eta}^{f\dagger}\big[\big(-\mathrm{i}\omega_{n}\big)\delta_{\eta\kappa}+{\cal{H}}_{\eta\kappa}^{f}\big]\Psi_{\kappa}^{f}+\cdots
\label{eq20}
\end{eqnarray}
Here $\Psi^{X}=\big(X_{\bm{k}}^{A}(\mathrm{i}v_{n}),X_{\bm{k}}^{B}(\mathrm{i}v_{n})\big)^{T}$ and $\Psi^{f}=\big(f_{\bm{k}\uparrow}^{A}(\mathrm{i}\omega_{n}),$
$f_{\bm{k}\uparrow}^{B}(\mathrm{i}\omega_{n}),f_{\bm{k}\downarrow}^{A}(\mathrm{i}\omega_{n}),f_{\bm{k}\downarrow}^{B}(\mathrm{i}\omega_{n}\big)^{T}$. 
Hamiltonian matrices of the X-field and $f$-field are respectively
\begin{eqnarray}
{\cal{H}}^{X}=
\begin{pmatrix}
\lambda Q_{X}^{\prime}\gamma_{X}&(-tQ_{X}+\alpha Q_{X}^{\prime\prime})g\\
(-tQ_{X}+\alpha Q_{X}^{\prime\prime})g^{*}&\lambda Q_{X}^{\prime}\gamma_{X}
\end{pmatrix}
\label{eq21}
\end{eqnarray}
and
\begin{widetext}
\begin{eqnarray}
{\cal{H}}^{f}=
\begin{pmatrix}
\lambda Q_{f}^{\prime}\gamma_{f}&-tQ_{f}g&0&\alpha Q_{f}(w_{x}-\mathrm{i}w_{y})\\
-tQ_{f}g^{*}&-\lambda Q_{f}^{\prime}\gamma_{f}&\alpha Q_{f}(w_{x}^{*}-\mathrm{i}w_{y}^{*})&0\\
0&\alpha Q_{f}(w_{x}+\mathrm{i}w_{y})&-\lambda Q_{f}^{\prime}\gamma_{f}&-tQ_{f}g\\
\alpha Q_{f}(w_{x}^{*}+\mathrm{i}w_{y}^{*})&0&-tQ_{f}g^{*}&\lambda Q_{f}^{\prime}\gamma_{f}
\end{pmatrix}.
\label{eq22}
\end{eqnarray}
\end{widetext}
In the Eq.(\ref{eq21}) and (\ref{eq22}), two new functions are $\gamma_{X}=2[\cos(\sqrt{3}ak_{y})+2\cos(3ak_{x}/2)\cos(\sqrt{3}ak_{y}/2)]$ and $\gamma_{f}=\gamma$. The expressions of $\gamma$, $g$, $w_{x}$ and $w_{y}$ are listed below the Eq.(\ref{eq3}).

\subsubsection{\label{sec3-1-2}Green's functions and self-consistency equations of two degrees of freedom}

The Green's function of the charge degree of freedom in the lower band (i.e. valence band) can be obtained from the action of Eq.(\ref{eq20}) using the standard formulae\cite{2015Coleman}. We get 
\begin{eqnarray}
G_{X}^{l}=\frac{1}{v_{n}^{2}/U+\rho+E_{X}^{l}}. 
\label{eq23}
\end{eqnarray}
Here $E_{X}^{l}=-|-tQ_{X}+\alpha Q_{X}^{\prime\prime}||g|+\lambda Q_{X}^{\prime}\gamma_{X}$ is the lower eigenenergy of the  Hamiltonian matrix  ${\cal{H}}^{X}$ of the X-field. Similarly, the Green's functions of the spinon in the two lower bands can be obtained as  
\begin{eqnarray}
G_{f}^{1(2)}=\frac{1}{\mathrm{i}\omega_{n}-E_{f}^{1(2)}}.
\label{eq24}
\end{eqnarray}
Here $E_{f}^{1}$ and $E_{f}^{2}$ ($E_{f}^{1}\leq E_{f}^{2}$) are the two lower eigenenergies of Hamiltonian matrix ${\cal{H}}^{f}$. Although it is actually not so easy to diagonalize the matrix ${\cal{H}}^{f}$ by hand, the eigenenergy $E_{f}$ and eigenvectors required for the mean field equations (see below) can be obtained more easily by computer both analytically and numerically. The pole of the Green's function in the imaginary frequency domain gives the energy spectrums of bands. Therefore, energy spectrums of the charge and spin degree of freedom (spinons) in the lower bands can be obtained respectively as 
\begin{eqnarray}
\xi^{l}(\bm{k})=\sqrt{U(\rho+E_{X}^{l})}
\label{eq25}
\end{eqnarray}and
\begin{eqnarray}
\Xi^{1(2)}(\bm{k})=E_{f}^{1(2)}.
\label{eq26}
\end{eqnarray}
Comparing Eq.(\ref{eq22}) with Eq.(\ref{eq3}), it is obvious that the energy spectrum $\Xi({\bm{k}})$ of the spinon in the case of interacting is quite the same as the one of the non-interacting electron, but the model parameters $t$, $\lambda$ and $\alpha$ are renormalized by the interactions. There are three definitions of these renormalized model parameters as follows  
\begin{equation}
t^{R}=Q_{f}t,\quad\lambda^{R}=Q_{f}^{\prime}\lambda\quad \text{and}\quad\alpha^{R}=Q_{f}\alpha.
\label{eq27}
\end{equation} 
The similarity between the two spectrums of spinons and non-interacing electron has important consequences for the Rashba SOC-driven topologically trivial and non-trivial transition that will be discussed later. 

The definitions of five mean field parameters, i.e. Eq.(\ref{eq10})--(\ref{eq14}) and the constraint equation of the X-field, i.e. $|X_{i}|^{2}=1$  are actually the six self-consistency equations in the slave-rotor mean field method. We obtain these self-consistency mean field equations as
\begin{eqnarray}
\frac{1}{2N_{\Lambda}}{\sum_{\bm{k}}}'\frac{\sqrt{U}}{2\sqrt{\rho+E_{X}^{l}}}+x_{0}=1,
\label{eq28}\\
Q_{f}=\frac{1}{6N_{\Lambda}}{\sum_{\bm{k}}}'\frac{\text{Sgn}(tQ_{X}-\alpha Q_{X}^{\prime\prime})|g|\sqrt{U}}{2\sqrt{\rho+E_{X}^{l}}}+x_{0},
\label{eq29}\\
Q_{f}^{\prime}=\frac{1}{12N_{\Lambda}}{\sum_{\bm{k}}}'\gamma_{X}\cdot\frac{\sqrt{U}}{2\sqrt{\rho+E_{X}^{l}}}+x_{0},
\label{eq30}
\end{eqnarray}
\begin{widetext}
\begin{eqnarray}
Q_{X}=\frac{1}{6N_{\Lambda}}\sum_{\bm{k}}[(u_{11}^{*}u_{12}+u_{21}^{*}u_{22}+u_{13}^{*}u_{14}+u_{23}^{*}u_{24})g+\text{c.c}],
\label{eq31}\\
Q_{X}^{\prime}=\frac{1}{12N_{\Lambda}}\sum_{\bm{k}}(|u_{11}|^{2}+|u_{14}|^{2}+|u_{21}|^{2}+|u_{24}|^{2}-|u_{12}|^{2}-|u_{13}|^{2}-|u_{22}|^{2}-|u_{23}|^{2})\gamma_{f},
\label{eq32}\\
Q_{X}^{\prime\prime}=\frac{1}{6N_{\Lambda}}\sum_{\bm{k}}[(w_{x}-\mathrm{i}w_{y})(u_{11}^{*}u_{14}+u_{21}^{*}u_{24})+(w_{x}^{*}-\mathrm{i}w_{y}^{*})(u_{12}^{*}u_{13}+u_{22}^{*}u_{23})+\text{c.c}].
\label{eq33}
\end{eqnarray}
\end{widetext}
Here $u_{11}$,$u_{12}$,$\cdots$, $u_{24}$ are the components of vectors $\bm{u}_{1}$ and $\bm{u}_{2}$. $\bm{u}_{1}=(u_{11},u_{12},u_{13},u_{14})^{T}$ and  $\bm{u}_{2}=(u_{21},u_{22},u_{23},u_{24})^{T}$ are the eigenvectors of the Hamiltonian matrix ${\cal{H}}^{f}$, corresponding to the two lower eigenenergies $E_{f}^{1}$ and $E_{f}^{2}$ ($E_{f}^{1}\leq E_{f}^{2}$) respectively and can be obtained analytically or numerically by computer.

\subsection{\label{sec3-2}The Mott transition of the charge degree of freedom}

\begin{figure}[ht]
\includegraphics[width=8.5cm,height=5.5cm]{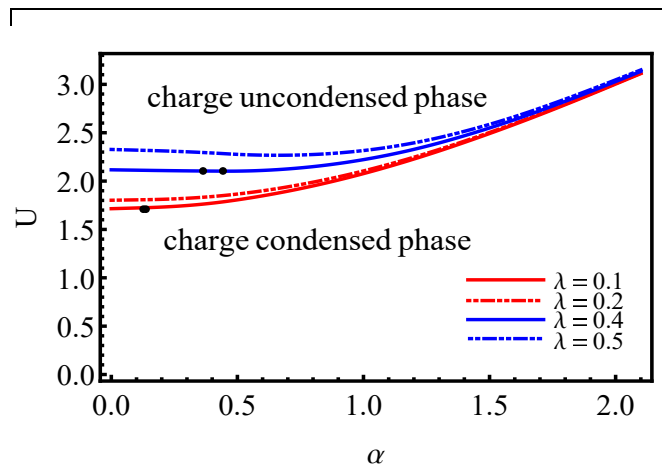}
\caption{\label{fig7}(Color online)Boundaries of the Mott transition of charge degree of freedom for various intrinsic SOC. The black points represent the special points on the boundaries of Mott transition at which the topologically trivial or non-trivial transition occurs (see section \ref{sec3-3}). The charge condensed and uncondensed phase are discussed in the main text.}  
\end{figure}

In the slave-rotor mean field method, the gap of the charge degree of freedom is closed when the interaction U is small. The condensed charge combines the spinon to form the conventional physical electron. In the larger-U region, the gap of the charge degree of freedom can be opened and a spin-charge separation occurs. There is a Mott transition of the charge degree of freedom. At the Mott transition, the density of the condensate of charges $x_0=0$ and $\rho=-\mathrm{min}(E_{X}^{l})$ derived from Eq.(\ref{eq25}). Under the two conditions, we can solve numerically the six mean field self-consistency equations i.e. Eq.(\ref{eq28})--(\ref{eq33}) and obtain the boundary of Mott transition as shown in Fig.\ref{fig7}.
From Eq.(\ref{eq3}) and Eq.(\ref{eq22}), the spinon has the same band structure as the one of conventional physical electron of the R-KM model which is topologically non-trivial in the case of small Rashba SOCs $\chi<\chi_{topo}$. Therefore, below the boundary of the Mott transition the combined phase may be a \textit{correlated} $Z_{2}$ topological insulator which connect adiabatically to the one possessed by the R-KM model. It have been investigated that the Rashba SOC can drive the R-KM model into a topologically non-trivial SM or trivial metal state from $Z_{2}$ topological insulators. Therefore, the charge condensed phase may also be a topologically non-trivial semi-metal or trivial metal state when the strength of the Rashba SOC beyond some critical values at which the indirect band gap of the spinon closes or the valence and conduction band can touch with each other. Above the boundary of the Mott transition, i.e. in the region of the charge uncondensed phase, it is a Mott insulator (MI) for the charge degree of freedom, while a quantum spin liquid (QSL) state for the spinon. There are also Rashba SOC-driven topologically trivial or non-trivial phase transitions of the spinon and novel quantum phases can emerge, as discussed below.

\subsection{\label{sec3-3}influences of correlation on Rashba SOC-driven phase transitions in the region of charge condensed phase}

\begin{figure}[ht]
\includegraphics[width=8.5cm,height=5.5cm]{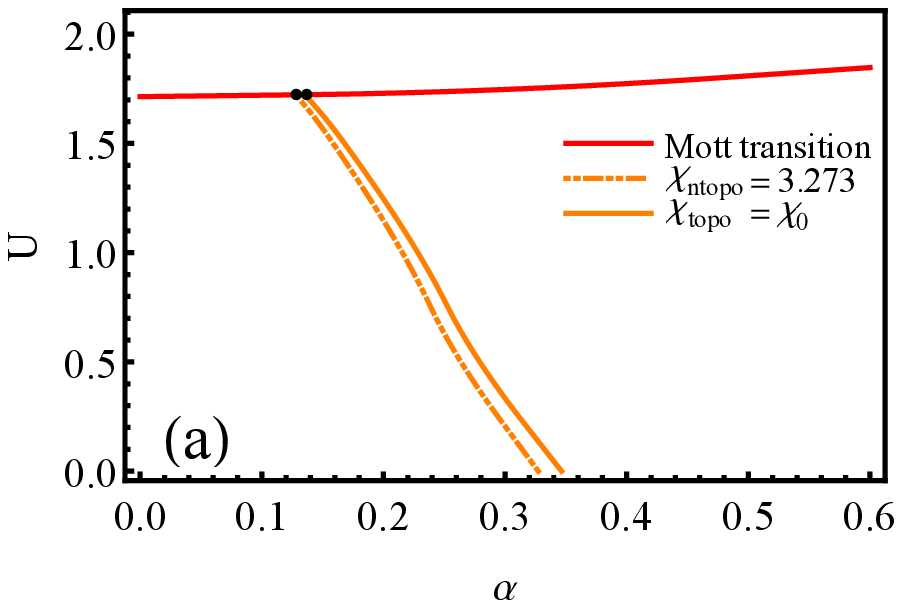}
\includegraphics[width=8.5cm,height=5.5cm]{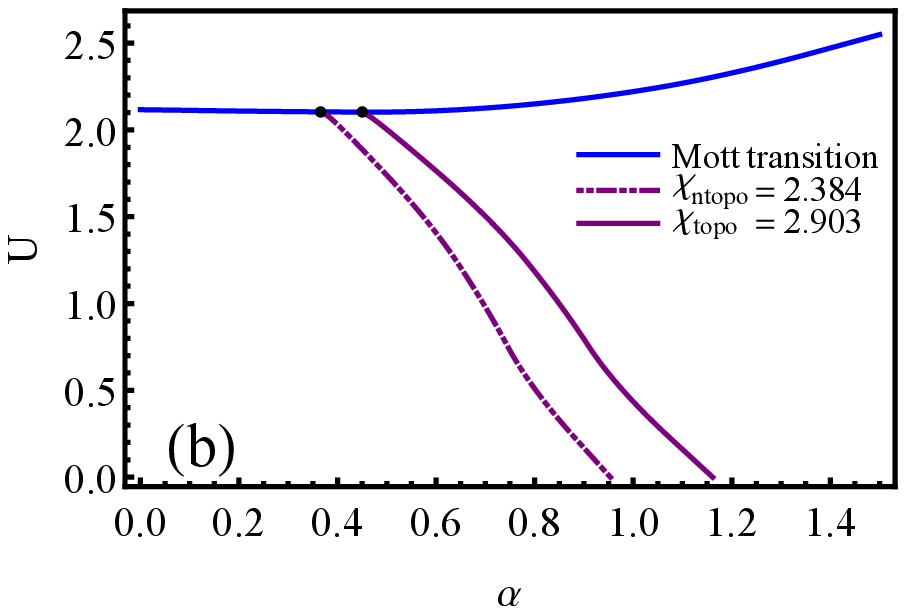}
\caption{\label{fig8}(Color online)Boundaries of topologically trivial and non-trivila transitions of the spinon below the Mott tansition of the charge degree of freedom. (a) $\lambda=0.1$, (b) $\lambda=0.4$. The black points represent the special points on the boundaries of Mott transition at which the topological trivial or non-trivial transition occurs (see main text).}  
\end{figure}

\begin{figure}[ht]
\includegraphics[width=8.5cm,height=5.5cm]{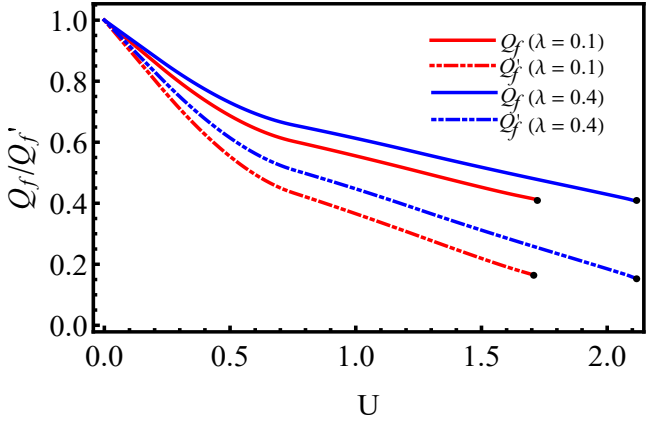}
\caption{\label{fig9}(Color online)Numerical results of $Q_{f}$ and $Q_{f}^{\prime}$ along the boundary of the topologically trivial transition. Black points represent the special points corresponding to Mott transition of charge degree of freedom. Along the boundary of topologically non-trivial transition $Q_{f}$ and $Q_{f}^{\prime}$ shift slightly and we don't draw them here.   }  
\end{figure}

As observed from Eq.(\ref{eq3}) and (\ref{eq22}), the energy band of the spinon has the same structure as the one of non-interacting electrons of R-KM model but with the renormalized electron hopping $t$, intrinsic SOC $\lambda$ and Rashba SOC $\alpha$, as defined in Eq.(\ref{eq27}). In the model without strong correlations, topologically trivial and non-trivial transitions of electrons occur at $\alpha=\chi_{ntopo}\cdot\lambda$ and $\alpha=\chi_{topo}\cdot\lambda$ respectively. It is obvious that, for spinons, the topologically trivial or non-trivial transition can occur at $\alpha^{R}=\chi_{ntopo}\cdot\lambda^{R}$ or $\alpha^{R}=\chi_{topo}\cdot\lambda^{R}$. Moreover, the density of the condensate of charges  $x_{0}\not=0$ at the point $(\mathrm{i}v_{n}^0,\bm{k}^0)$ and the gap of the charge degree of freedom is closed, i.e. $\rho=-\mathrm{min}(E_{X}^{l})$ when charges condense. Thus, we obtain the conditions 
\begin{equation}
x_{0}\not=0,\quad\rho=-\mathrm{min}(E_{X}^{l}),\quad\alpha=\chi_{ntopo}\cdot\lambda\frac{Q_{f}^\prime}{Q_{f}}
\label{eq34}
\end{equation}
for topologically trivial transition of spinons and 
\begin{equation}
x_{0}\not=0,\quad\rho=-\mathrm{min}(E_{X}^{l}),\quad\alpha=\chi_{topo}\cdot\lambda\frac{Q_{f}^\prime}{Q_{f}}
\label{eq35}
\end{equation}
for topologically non-trivial transition of spinons in the region of the charge condensed phase. The Rashba SOC-driven topologically trivial and non-trivial transition in the charge condensed phase can be obtained by solving numerically the self-consistency equations (\ref{eq28})--(\ref{eq33}) under the conditions (\ref{eq34}) and (\ref{eq35}) respectively. The results are shown in Fig.\ref{fig8}.  
For $U=0$, we reproduce the earlier result of non-interacting limit that is $\alpha=\chi_{ntopo}\cdot\lambda$ and $\alpha=\chi_{topo}\cdot\lambda$ for the topologically trivial and non-trivial transition respectively, because of $Q_{f}=1$ and $Q_{f}^{\prime}=1$ at $U=0$. The behavior of $Q_{f}$ and $Q_{f}^{\prime}$ is shown in Fig.\ref{fig9}.

An important investigation we want to make is the influence of strong correlations on the Rashba SOC-driven transition. From our numerical results, it can be seen that the critical $\alpha$ for both topologically trivial and non-trivial transition shifts to the smaller value with increasing $U$ for each $\lambda$. There is a narrow window where a topological SM exists. It is similar to the case of non-interacting limit. For topologically trivial transition, the region of correlated $Z_{2}$ topological insulator is shrank and the correlation destabilize the correlated topological phase.  The correlation has the similar behavior on the topologically non-trivial transition, i.e. it  destabilize the correlated topological SM phase. As a consequence, the region of topological SM phase shrinks slightly with the increasing correlation. Furthermore, the larger intrinsic SOC can lead to the wider region of topological SM phase. For the small intrinsic SOC the region is very narrow, e.g. the case of $\lambda=0.1$ as shown in Fig.\ref{fig8}(a). 

In particular, when the phase boundary of spinon reaches the Mott boundary, the topologically trivial or non-trivial transition of spinons and the Mott transition of charge degree of freedom occur simultaneously. This can also be obtained from self-consistency equations under the condition as $x_{0}=0,\rho=-\mathrm{min}(E_{X}^{l})$ and $\alpha=\chi_{(n)topo}\cdot\lambda\frac{Q_{f}^\prime}{Q_{f}}$ and the special transition points are marked out by black points in Fig.\ref{fig7} and \ref{fig8}. 
      
\begin{figure}[ht]
\includegraphics[width=8.5cm,height=5.5cm]{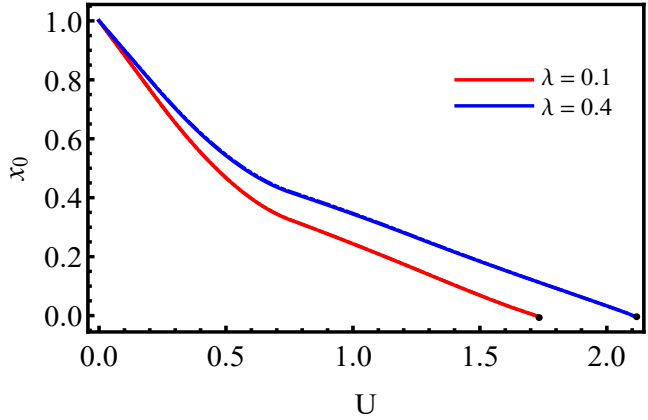}
\caption{\label{fig10}(Color online)The decreasing charge condensate density $x_{0}$ with the increasing correlation along the boundary of topologically trivial transition of spinons. Black points represent the special points corresponding to Mott transition of charge degree of freedom.}  
\end{figure}

Along the toplogically trivial or non-trivial phase boundary, the condensate density $x_{0}$ should decrease with the increasing correlation. The behavior of the condensate density is shown in Fig.\ref{fig10}. The condensate density has the maximum value $x_{0}=1$ at $U=0$ and is equal to zero when the phase boundary reaches the Mott boundary. In our numerical calculation, for each intrinsic SOC, the condensate density along the boundary of the topologically non-trivial transition shifts slightly compared to the one along the boundary of the topologically trivial transition. So we don't draw it here.

\subsection{\label{sec3-4}influences of correlation on Rashba SOC-driven phase transitions in the charge Mott region}

\begin{figure}[ht]
\includegraphics[width=8.5cm,height=5.5cm]{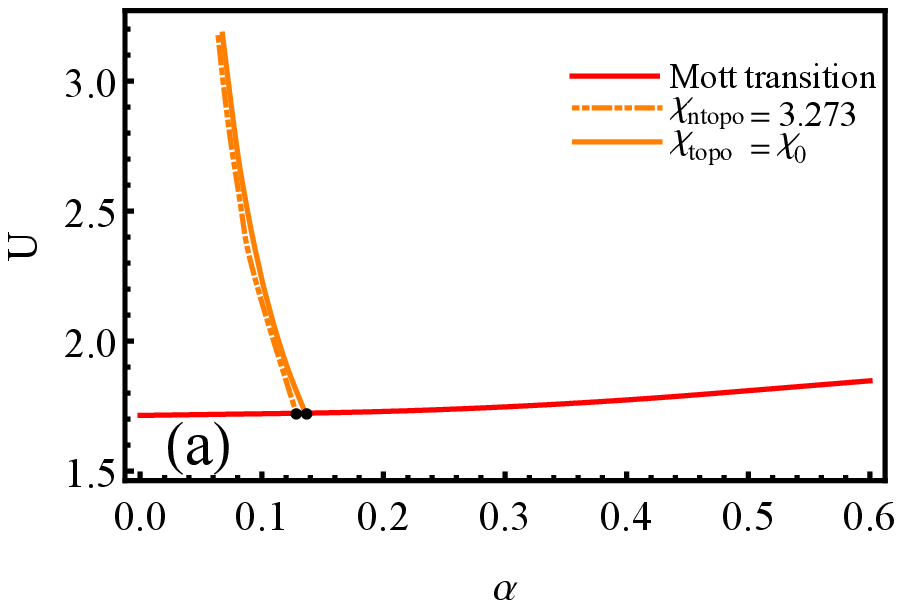}
\includegraphics[width=8.5cm,height=5.5cm]{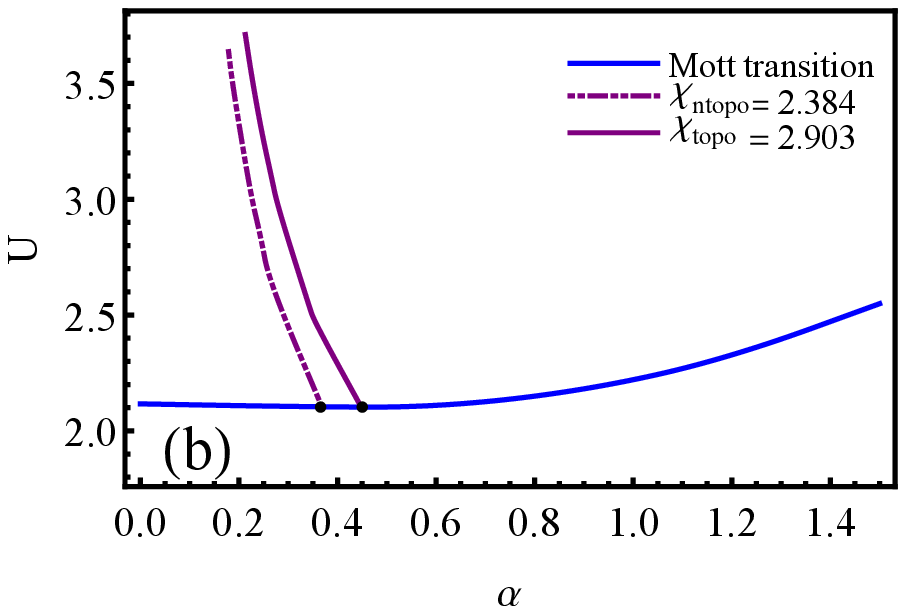}
\caption{\label{fig11}(Color online)Boundaries of topologically trivial and non-trivila transitions of the spinon in the Mott region of the charge degree of freedom. (a) $\lambda=0.1$, (b) $\lambda=0.4$. The black points represent the special points on the boundaries of Mott transition at which the topological trivial or non-trivial transition occurs.}  
\end{figure}

In the Mott region of the intermediate strength of correlations, there is no condensate of the charge degree of freedom, i.e. $x_{0}=0$ and the gap of charge degree of freedom is opened, i.e. $\rho\not=-\mathrm{min}(E_{-}^{X})$. The conditions that the topologically trivial and non-trivial transition occur in the Mott region of charge degree of freedom can be obtained respectively as
\begin{equation}
x_{0}=0,\quad\rho\not=-\mathrm{min}(E_{X}^{l}),\quad\alpha=\chi_{ntopo}\cdot\lambda\frac{Q_{f}^\prime}{Q_{f}},
\label{eq36}
\end{equation} 
and
\begin{equation}
x_{0}=0,\quad\rho\not=-\mathrm{min}(E_{X}^{l}),\quad\alpha=\chi_{topo}\cdot\lambda\frac{Q_{f}^\prime}{Q_{f}}
\label{eq37}.
\end{equation} 
Under the condition (\ref{eq36}) or (\ref{eq37}) for the Rashba SOC-driven transition, the altered self-consistency equations are solvable. Boundaries of the topologically trivial and non-trivial transition in the Mott region of charge degree of freedom are shown in Fig.\ref{fig11}.

The influences of correlations on the Rashab SOC-driven topologically trivial or non-trivial transition in the charge Mott region are similar to the case of condensed charges. For topologically trivial transition, the correlation destabilizes the insulating phase of the spinon which possesses the same  band structure as the $Z_{2}$ topological insulator in the region of condensed charges. Before the topologically trivial transition, the phase is a mixed state of the Mott insulator of charges and the QSL of spinons with the topologically non-trivial band structure, which is actually a topological Mott insulator (TMI)\cite{2010Rachel,2010Pesin}. After the topologically trivial transition, the indirect energy gap of spinons is closed and the mixed state becomes a TMI with gapless spin excitations (we call it TMI$^{*}$).  For the topologically non-trivial transition, the correlation also destabilizes the topologically non-trivial phase (TMI$^{*}$). After this transition, the QSL component in the mixed phase becomes a topologically trivial state due to the band touching of valence and conduction bands of spinons, but maintains the gapless spin excitation. 

To consider the stability of the QSLs or TMI and TMI$^{*}$ associated with quantum fluctuations, our zeroth-order mean-field method should be extented to the first-order mean-field theory via the introduction of the U(1) or SU(2) gauge field coupled to the spinon\cite{2004Wen}. To obtain a stable mean-field QSL or TMI (or TMI$^{*}$), we should give gauge fluctuations a finite energy gap. Young \textit{et al.}\cite{2008Young} have supposed that when another coupled honeycomb lattice layer is added one can open up a gap for gauge field. This suppresses the gauge fluctuation, and then the QSLs or TMI and TMI$^{*}$ are stable in this situation. So we are not concerned with quantum fluctuations and assume that these mean-field QSLs are stable in this work.

\begin{figure}[ht]
\includegraphics[width=8.5cm,height=6.0cm]{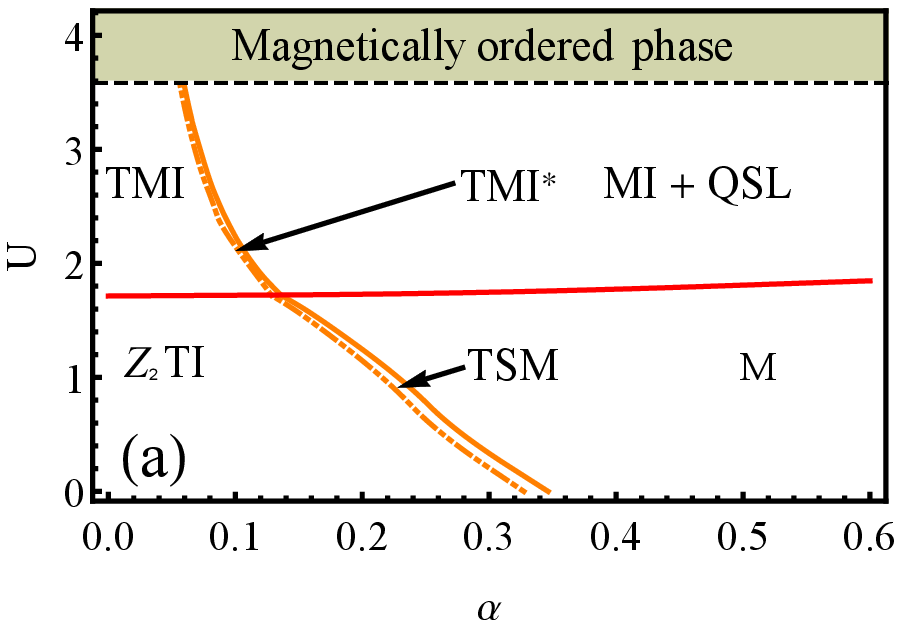}
\includegraphics[width=8.5cm,height=6.0cm]{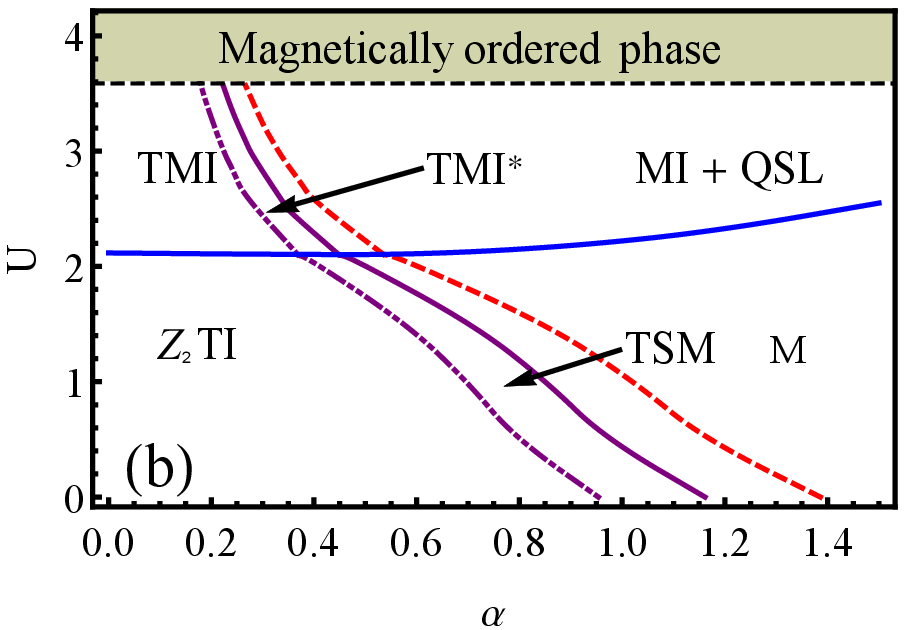}
\caption{\label{fig12}(Color online)Phase diagrams of the R-KMH model at (a) $\lambda=0.1$ and (b) $\lambda=0.4$. The magnetically ordered phase at the larger U have been discussed in Ref.[\onlinecite{2014Laubach}]. The red dashed line in (b) corresponds to the $\chi_{0}$($=2\sqrt{3})$ and does not the boundary of topologically trivial or non-trivial transition at $\lambda=0.4$.}  
\end{figure}

We can summarize above discussions to obtain the phase diagram of the R-KMH model as shown in Fig.~\ref{fig12}. 

\section{\label{sec4}conclusions and outlook}

The Rashba SOC has dramatic effects on the original KM model which possesses the intrinsic SOC. It can lift the spin degeneracy of bands and break the mirror symmetry of KM model. Besides, the competition between the Rashba and intrinsic SOCs can lead to the band touching or the direct gap closing. For the small intrinsic SOC, the band touching occurs at $\chi_{topo}=\chi_{0}=2\sqrt{3}$ that is independent of the intrinsic SOC. This is the classical result at which topologically non-trivial transition occurs. When $\chi>\chi_{topo}$, the Rashba  SOC can't open the gap, but maintains touching points at $K$ and $K^{\prime}$ and adds three satellite touching point around the $K$ or $K^{\prime}$. For the large intrinsic SOC, the critical $\chi_{topo}$ at which topologically non-trivial transition or the band touching occurs decreases with the increasing intrinsic SOC due to the deformation of the band caused by the Rashba SOC. There are three touching points which are located around $K$ and $K^{\prime}$. When $\chi_{topo}<\chi<\chi_{0}$, the three points are split into six touching points. The inside three points move towards the $K$ or $K^{\prime}$ and eventually shrinks into a single point at $K$ or $K^{\prime}$ when $\chi=\chi_{0}$. The phenomenon when $\chi>\chi_{0}$ is the same as the one when $\chi>\chi_{topo}$ in the case of small intrinsic SOCs. Furthermore, there is an indirect gap closing before the band touching occurs for all the intrinsic SOC where the topologically tivial transition occurs. The two Rashba SOC-driven transitions lead to three distinct phases possessed by R-KM model, i.e. the $Z_{2}$ TI, the TSM phase and the Metal phase.

We investigate effects of correlation using slave-rotor mean field method. There is a Mott transition of charge degree of freedom and the band topology of the electron in the non-interacting model is inherited by the spinon. Below the Mott boundary, the condensed charge combines spinon to form the physical electrons. Correlations can let the smaller Rashba SOC to drive the topologically trivial and non-trivial transition of spinons (also physical electrons) and destabilize correlated $Z_{2}$ TI and TSM phase. Above the Mott transition, the correlation has the similar behavior on the two Rashba SOC-driven transition of spinons. Because of the uncondensed charge in the Mott region, the correlation destabilizes two topological mixed phases of the charge degree of freedom and the spinon (not the physical electron now), i.e. TMI and TMI$^{*}$. The effects of correlations are summarized in phase diagrams of the R-KMH model (see Fig.(\ref{fig12})).

The breaking of both mirror and hexagonal symmetries can lead to the anisotropic Rashba SOC, effects of which on the energy spectrum of the two-dimensional electron gas\cite{2016Johansson}, metallic surface states\cite{2010Simon} or electrons in graphene sheet\cite{2017Hosseini} have been  investigated. The spin polarization can be influenced by the anisotropic Rashba SOC and there is a Lifshitz transition. The effects of the anisotropy of the Rahsba SOC may make the competition between the Rashba SOC and the intrinsic SOC more interesting. Furthermore, the correlation may also have important influences on the competition in the case of the anisotropic Rashba SOC.   
    
\begin{acknowledgments}
This work was supported by National Natural Science Foundation of China under Grant No. 11964042 and the Scientific Research Project of the Education Department of Sichuan Province under Grant No. 18ZB0464.
\end{acknowledgments}


\begin{thebibliography}{99}                                                                                             
%
\bibitem {2010Hasan}M. Z. Hasan and C. L. Kane, Rev. Mod. Phys. 82, 3045 (2010).
\bibitem {2011Qi}X. L. Qi and S. C. Zhang, Rev. Mod. Phys. 83, 1075 (2011).
\bibitem {2010Xiao}D. Xiao, M. C. Chang and Q. Niu, Phys. Mod. Phys. 82, 1959 (2010).
\bibitem {2010Wang}Z. Wang, X. L. Qi and S. C. Zhang, New J. Phys. 12, 065007 (2010).
\bibitem {2007Moore}J. E. Moore and L. Balents, Phys. Rev. B 75, 121306(R) (2007)
\bibitem {2003Avron}J. A. Avron, Phys. Today 56, 38 (2003).
\bibitem {2005aKane}C. L. Kane and E. J. Mele, Phys. Rev. Lett. 95, 226801 (2005).
\bibitem {2005bKane}C. L. Kane and E. J. Mele, Phys. Rev. Lett. 95, 146802 (2005).
\bibitem {1988Haldane}F. D. M. Haldane, Phys. Rev. Lett. 61, 2015 (1988).
\bibitem {1985Kohmoto}M. Kohmoto, Ann. Phys. 160, 343 (1985).
\bibitem {2007Yao}Y. Yao, F. Ye, X. L. Qi, S. C. Zhang and Z. Fang, Phys. Rev. B 75, 041401(R) (2007).
\bibitem {2006Bernevig}B. A. Bernevig, T. L. Hughes and S. C. Zhang, Science 314, 1757 (2006).
\bibitem {2007Konig}M. K\"onig, S. Wiedmann, C. Br\"une, A. Roth, H. Buhmann, L. Molenkamp, X. L. Qi and S. C. Zhang, Science 318, 766 (2007).
\bibitem {1960Rashba}E. I. Rashba, Sov. Phys. Solid State 2, 1109 (1960).
\bibitem {1984Bychkov}Yu. A. Bychkov and E. I. Rashba, JETP Lett. 39, 78 (1984).
\bibitem {2008Varyhalov}A. Varykhalov, J. Sanchez-Barriga, A. M. Shikin, C. Biswas, E. Vescovo, A. Rybkin, D. Marchenko and O. Rader, Phys. Rev. Lett. 101, 157601 (2008).
\bibitem {2009Castro Neto}A. H. Castro Neto and F. Guinea, Phys. Rev. Lett. 103, 026804 (2009).
\bibitem {2011Weeks}C. Weeks, J. Hu, J. Alicea, M. Franz and R. Wu, Phys. Rev. X 1, 021001 (2011).
\bibitem {2009Zarea}M. Zarea and N. Sandler, Phys. Rev. B 79,165442 (2009).
\bibitem {2009Rashba}E. I. Rashba, Phys. Rev. B 79, 161409(R) (2009).
\bibitem {2013Shakouri}Kh. Shakouri, M. Ramezani Masir, A. Jellal, E. B. Choubabi and F. M. Peeters, Phys. Rev. B 88,115408 (2013).
\bibitem {2005Sheng}L. Sheng, D. N. Sheng, C. S. Ting and F. D. M. Haldane, Phys. Rev. Lett. 95, 136602 (2005).
\bibitem {2006Sheng}D. N. Sheng, Z. Y. Weng, L. Sheng and F. D. M. Haldane, Phys. Rev. Lett. 97, 036808 (2006).
\bibitem {1984Niu}Q. Niu and D. J. Thouless, J. Phys. A: Math. Gen 17, 2453 (1984).
\bibitem {1985Niu}Q. Niu, D. J. Thouless and Y. S. Wu, Phys. Rev. B 31, 3372 (1985).
\bibitem {2008Young}M. W. Young, S. S. Lee and C. Kallin, Phys. Rev. B 78, 125316(2008).
\bibitem {2008Cai}Z. Cai, S. Chen, S. P. Kou and Y. P. Wang, Phys. Rev. B 78, 035123 (2008).
\bibitem {2013Hohenadler}M. Hohenadler and  F. F. Assaad,  J. Phys.: Condens. Matter 25, 143201 (2013).
\bibitem {2014Witczak-Krempa}W. Witczak-Krempa, G. Chen, Y. B. Kim and L. Balents, Ann. Rev. Condens. Matter Phys. 5, 57 (2014).
\bibitem {2018Rachel}S. Rachel, Rep. Prog. Phys. 81, 116501 (2018).
\bibitem {2010Rachel}S. Rachel and K. Le Hur, Phys. Rev. B 82, 075106 (2010).
\bibitem {2012Ruegg}A. R\"uegg  and G. A. Fiete, Phys. Rev. Lett. 108, 046401 (2012).
\bibitem {2012Vaezi}A. Vaezi, M. Mashkoori and M. Hosseini, Phys. Rev. B 85, 195126 (2012).
\bibitem {2012Wu}W. Wu, S. Rachel, W. M. Liu and K. Le Hur, Phys. Rev. B 85, 205102 (2012).
\bibitem {2011Yu}S. L. Yu, X. C. Xie and J. X. Li, Phys. Rev. Lett. 107, 010401 (2011).
\bibitem {2011Hohenadler}M. Hohenadler, T. C. Lang and F. F. Assaad, Phys. Rev. Lett. 106, 100403 (2011).
\bibitem {2011Zheng}D. Zheng, G. M. Zhang and C. Wu, Phys. Rev. B 84, 205121 (2011).
\bibitem {2012Hohenadler}M. Hohenadler, Z. Y. Meng, T. C. Lang, S. Wessel, A. Muramatsu and F. F. Assaad, Phys. Rev. B 85, 115132 (2012).
\bibitem {2012Griset}C. Griset and C. Xu, Phys. Rev. B 85, 045123 (2012). 
\bibitem {2014Bercx}M. Bercx, M. Hohenadler and F. F. Assaad, Phys. Rev. B 90, 075140 (2014).
\bibitem {2010Strom}A. Str\"om, H. Johannesson and G. I. Japaridze, Phys. Rev. Lett. 104, 256804 (2010).
\bibitem {2012Schmidt}T. L. Schmidt, S. Rachel, F. von Oppen and L. I. Glazman, Phys. Rev. Lett. 108,156402 (2012).
\bibitem {2012Budich}J. C. Budich, F. Dolcini, P. Recher and B. Trauzettel, Phys. Rev. Lett. 108,086602 (2012)
\bibitem {2014Hohenadler}M. Hodenadler and F. F. Assaad, Phys. Rev. B 90, 245148 (2014).
\bibitem {2014Laubach}M. Laubach, J. Reuther, R. Thomale and S. Rachel, Phys. Rev. B 90, 165136 (2014).
\bibitem {2018Mishra}A. Mishra and S. Lee, Sci. Rep. 8,799 (2018).
\bibitem {1997Gebhard}F. Gebhard, \textit{The mott metal-insulator transition: models and methods} (Springer-Verlag, Berlin Heidelberg, 1997)
\bibitem {2002Florens}S. Florens and A. Georges, Phys. Rev. B 66, 165111 (2002).
\bibitem {2003Florens}S. Florens, P. San Jos\'e, F. Guinea and  A. Georges, Phys. Rev. B 68, 245311 (2003).
\bibitem {2004Florens}S. Florens and A. Georges, Phys. Rev. B 70, 035114 (2004). 
\bibitem {2015Coleman}P. Coleman, \textit{Introduction to Many-Body Physics} (Cambridge University Press, Cambridge, 2015)
\bibitem {2010Pesin} D. A. Pesin and L. Balents, Nat. Phys. 6, 376 (2010).
\bibitem {2004Wen}X. G. Wen, \textit{Quantum Field Theory of Many-Body Systems} (Oxford University Press, New York, 2004)
\bibitem {2016Johansson}A. Johansson, J. Henk and I. Mertig, Phys. Rev. B 93, 195440 (2016).
\bibitem {2010Simon}E. Simon, A. Szilva, B. Ujfalussy, B. Lazarovits, G. Zarand and L. Szunyogh, Phys. Rev. B 81, 235438 (2010).
\bibitem {2017Hosseini}M. V. Hosseini, J. Phys.: Condens. Matter 29, 315502 (2017)

\end{thebibliography}
\end{document}